\documentclass[twocolumn, tighten, times, twocolappendix]{aastex631}

\begin{document}

\title{Realistic predictions for Gaia black hole discoveries: comparison of isolated binary and dynamical formation models}

\author[0000-0002-1386-0603]{Pranav Nagarajan}
\affiliation{Department of Astronomy, California Institute of Technology, 1200 E. California Blvd., Pasadena, CA 91125, USA}

\author[0000-0002-6871-1752]{Kareem El-Badry}
\affiliation{Department of Astronomy, California Institute of Technology, 1200 E. California Blvd., Pasadena, CA 91125, USA}

\author[0000-0001-9685-3777]{Chirag Chawla}
\affiliation{Department of Astronomy and Astrophysics, Tata Institute of Fundamental Research, Homi Bhabha Road, Navy Nagar, Colaba, Mumbai, 400005, India}

\author[0000-0003-2654-5239]{Ugo Niccol\`o Di Carlo}
\affiliation{Scuola Internazionale Superiore di Studi Avanzati (SISSA), Via Bonomea 265, I-34136 Trieste, Italy}

\author[0000-0001-5228-6598]{Katelyn Breivik}
\affiliation{McWilliams Center for Cosmology, Department of Physics, Carnegie Mellon University, 5000 Forbes Avenue, Pittsburgh, PA 15213, USA}

\author[0000-0003-4175-8881]{Carl L. Rodriguez}
\affiliation{Department of Physics and Astronomy, University of North Carolina at Chapel Hill, 120 E. Cameron Avenue, Chapel Hill, NC 27599, USA}

\author[0000-0002-1135-984X]{Poojan Agrawal}
\affiliation{Institute of Astronomy, KU Leuven, Celestijnenlaan 200D, B-3001, Leuven, Belgium}

\author[0000-0001-7099-765X]{Vera Delfavero}
\affiliation{Gravitational Astrophysics Laboratory, NASA Goddard Space Flight Center, Greenbelt, MD 20771, USA}

\author[0000-0002-3680-2684]{Sourav Chatterjee}
\affiliation{Department of Astronomy and Astrophysics, Tata Institute of Fundamental Research, Homi Bhabha Road, Navy Nagar, Colaba, Mumbai, 400005, India}



\begin{abstract}

Astrometry from {\it Gaia} has enabled discovery of three dormant black holes (BHs) in au-scale binaries. Numerous models have been proposed to explain their formation, including several that have forecasted {\it Gaia} detections. However, previous works have used simplified detectability metrics that do not capture key elements of the {\it Gaia} astrometric orbit selection function. We apply a realistic forward-model of {\it Gaia} astrometric orbit catalogs to BH binary populations generated through (a) isolated binary evolution (IBE) and (b) dynamical formation in star clusters. For both formation channels, we analyze binary populations in a simulated Milky Way-like galaxy with a realistic metallicity-dependent star formation history and 3D dust map. We generate epoch astrometry for each binary from the {\it Gaia} scanning law and fit it with the cascade of astrometric models used in {\it Gaia} DR3. The IBE model of \citet{chawla_gaia_2022} predicts that no BH binaries should have been detected in DR3 and thus significantly underpredicts the formation rate of Gaia BHs. In contrast, the dynamical model of \citet{di_carlo_young_2024} overpredicts the number of BHs receiving DR3 orbital solutions by a factor of $\sim8$. The two models predict very different orbital period distributions, with the IBE model predicting only binaries that avoided common envelope evolution and have $P_{\text{orb}} \gtrsim 2,000$ d to be detectable, and the dynamical formation model predicting a period distribution that is roughly log-uniform. Adopting the dynamical channel as a fiducial model and rescaling by a factor of $1/8$ to match DR3, we predict that $\sim30$ BH binaries will be detected in {\it Gaia} DR4, representing $\sim0.1\%$ of Milky Way BHs with luminous companions in au-scale orbits.


\end{abstract}



\section{Introduction} \label{sec:intro}

Astrometry from the third data release of the \textit{Gaia} mission \citep[DR3;][]{gaia_collaboration_gaia_2016, gaia_collaboration_gaia_2023, gaia_collaboration_gaia_2023-1} has allowed for the discovery and characterization of two dormant stellar-mass black holes (BHs) \citep{el-badry_sun-like_2023, chakrabarti_noninteracting_2023, nagarajan_espresso_2024, tanikawa_2023, el-badry_red_2023}. In DR3, epoch-level astrometric data was not published, and several conservative quality cuts were applied to the published orbital solutions. More BHs are expected to be discovered in future data releases. Indeed, a $33\,M_{\odot}$ BH orbited by a metal-poor red giant has already been discovered in \textit{Gaia} DR4 pre-release astrometry \citep{gaia_collaboration_discovery_2024}.

The formation of Gaia BH binaries remains uncertain. Isolated binary evolution (IBE) models struggle to reproduce the systems' wide orbits. Their current orbital separations are significantly smaller than the predicted maximum radii of their BH progenitors, suggesting that their luminous stars would have interacted with the progenitors when they were red supergiants. However, these binaries would also have to survive an episode of common-envelope evolution with a large donor-to-accretor mass ratio; even if the envelopes of the massive stars were successfully ejected, the final orbits are predicted to be far tighter than observed. Modified prescriptions of IBE \citep[e.g.,][]{kruckow_2024, gilkis_2024, kotko_2024, iorio_2024}, formation in a hierarchical triple \citep[e.g.,][]{nagarajan_espresso_2024, generozov_perets_2024, li_2024, tanikawa_bbh_2024} and dynamical assembly in a star cluster \citep[e.g.,][]{rastello_2023, tanikawa_2024, di_carlo_young_2024, fantoccoli_2024, pina_2024} have all been proposed as alternative explanations, but we cannot yet tell which of these models are correct, or their relative contributions.

Numerous studies published prior to the release of \textit{Gaia} DR3 have attempted to estimate the number of BHs that will be discovered by \textit{Gaia} \citep[e.g.,][]{mashian_hunting_2017, breivik_revealing_2017, yamaguchi_2018, yalinewich_2018, shao_li_2019, andrews_weighing_2019, shikauchi_2020, wiktorowicz_noninteracting_2020, chawla_gaia_2022, janssens_uncovering_2022, shikauchi_detectability_2022, wang_2022}. Most of these papers only considered the IBE formation channel, and predictions for the number of \textit{Gaia}-detectable BHs have generally become more pessimistic over time \citep[e.g.,][]{el-badry_gaias_2024}. So far, the \textit{Gaia} mission has found far fewer BHs than predicted in any of these works. Predictions published after \textit{Gaia} DR3 have mainly focused on dynamical formation scenarios, which can potentially avoid the challenges associated with the IBE channel \citep[e.g.,][]{tanikawa_2024, di_carlo_young_2024}. 

Studies so far have relied on simplified detectability estimates, rather than a realistic forward-model of the \textit{Gaia} astrometric binary catalogs. This approach is unlikely to yield accurate predictions because the selection function for orbital solutions published in DR3 is complex. For example, many sources that could have received well-constrained orbital solutions were instead published with acceleration solutions \citep{Pourbaix2022, 2024OJAp....7E.100E}. Also, the cuts used in DR3 depend on quantities that are nontrivial to predict with simplified detectability metrics, such as the uncertainties on astrometric parameters \citep{halbwachs_gaia_2023}. Recently, \citet{2024OJAp....7E.100E} built a forward model to generate synthetic astrometric binary catalogs for DR3 and future \textit{Gaia} data releases from the \textit{Gaia} scanning law. In this paper, we apply this generative model to simulated populations of BH-luminous companion (BH-LC) binaries in the Milky Way to predict the number of BHs detectable in \textit{Gaia} DR3, DR4, and DR5. Specifically, we investigate synthetic populations that assume either IBE \citep{chawla_gaia_2022} or dynamical \citep{di_carlo_young_2024} formation channels for Gaia BHs. By utilizing the generative model of \citet{2024OJAp....7E.100E} and calibrating with the number of Gaia BHs detected in DR3, we are able to make the most realistic predictions yet for the number of Gaia BHs that will be detected in DR4 and DR5.

We organize the remainder of this paper as follows. In Section~\ref{sec:populations}, we describe the synthetic BH-LC populations formed via IBE or dynamical assembly. In Section~\ref{sec:results}, we analyze the observational predictions of each of these formation channels by generating mock astrometric binary catalogs based on a realistic description of \textit{Gaia}'s astrometric model cascade. In Section~\ref{sec:discussion}, we discuss the implications of our results for future discoveries of dormant BHs in binaries. Finally, in Section~\ref{sec:conclusion}, we summarize our key results and provide pathways for future investigation.

\section{Population Synthesis} \label{sec:populations}

We consider two population synthesis models: one based on IBE alone, and one combining binary evolution and cluster dynamics. Both models are built on the same simulated Milky Way-like galaxy from the FIRE-2 simulation suite \citep{wetzel_2016, hopkins_2018}.

\subsection{Isolated Binary Evolution}
\label{sec:chawla}

For the isolated binary formation model, we adopt the population of BH-LC systems synthesized by \citet{chawla_gaia_2022}. These binaries are evolved from ZAMS to present day using \texttt{COSMIC} \citep{breivik_cosmic_2020}, which is based on a modified version of the binary-star evolution code \texttt{BSE} \citep{hurley_2002}. \citet{chawla_gaia_2022} consider both the ``rapid'' and ``delayed'' prescriptions for BH masses and natal kicks presented in \citet{fryer_2012}; for sake of comparison with \citet{di_carlo_young_2024}, we consider the population produced using the ``rapid'' mechanism. We combine the BH-main sequence (MS) and BH-post-main sequence (PMS) populations. 

In their initialization, \citet{chawla_gaia_2022} assign ages and metallicities for each binary based on the simulated galaxy m12i from the FIRE-2 simulation suite \citep{wetzel_2016, hopkins_2018}, and assign 3D Galactic positions based on the \textsf{ananke} framework of \citet{sanderson_2020}, which generates mock catalogs of synthetic stars from cosmological baryonic simulations. The metallicities are constrained to fall within the valid range of $\log(Z / Z_{\odot}) \in [-2.3, 0.2]$. Zero-age main sequence (ZAMS) orbital parameters are drawn from observationally motivated probability distribution functions; see \citet{chawla_gaia_2022} for details. \citet{chawla_gaia_2022} use the $\alpha \lambda$ prescription for common-envelope evolution, setting $\alpha = 1$ and the value of $\lambda$ based on the default prescription in the Appendix of \citet{claeys_2014}. They assume that BH natal kicks are reduced by a factor of $1 - f_{\text{fb}}$ relative to velocities drawn from a Maxwellian distribution with $\sigma = 265$ km s$^{-1}$ \citep{hobbs_statistical_2005}, with $f_{\text{fb}}$ representing the fractional fallback of the stellar envelope set by the prescription of \citet{fryer_2012}. 

To generate synthetic Milky Way (MW) BH-LC populations, \citet{chawla_gaia_2022} sample with replacement from a converged population of simulated binaries, where convergence refers to an iterative process in which the shapes of binary parameter distributions stabilize as the simulated population grows \citep{breivik_cosmic_2020}. To determine the number of BH-LC systems in the present-day MW, they scale up the number of BH-LC systems in the converged population by the ratio of the total mass formed in the galaxy m12i to the total mass of single and binary stars drawn to produce the simulated BH-LC population. The sampled binaries are assigned Galactocentric positions by minimizing the difference in age and metallicity with star particles in m12i; if multiple BH–LCs are assigned to the same single-star particle, they are distributed in a sphere centered around the particle with radius inversely dependent on the local density. 

\subsection{Binary Evolution and Cluster Dynamics}
\label{sec:dicarlo}

For the dynamical formation model, we adopt the population of BH-LC systems synthesized by \citet{di_carlo_young_2024}. In their 100-Myr long simulations of 3000 open star clusters \citep[``set A'', see][]{di_carlo_2020}, dynamics are treated by the $N$-body code \texttt{NBODY6++GPU} \citep{wang_2015}, and population synthesis by \texttt{MOBSE} \citep{mapelli_2017, giacobbo_2018}. The star clusters (SCs) have masses ranging from $10^3$--$3 \times 10^4\,M_{\odot}$ (drawn from the $dN/dM_{\text{SC}} \propto M_{\text{SC}}^{-2}$ young SC initial mass function (IMF) of \citealt{lada_2003}) and initial half-mass radii calculated using the relationship from \citet{marks_kroupa_2012}. The SCs have three different metallicities $Z = 0.02, 0.002,$ and $0.0002$.

To generate the SC initial conditions, \citet{di_carlo_young_2024} utilize \texttt{MCLUSTER} \citep{kupper_2011}. They adopt a \citet{kroupa_2001} IMF and draw binary parameters from the distributions of \citet{sana_binary_2012}. To calculate BH masses, they use the ``rapid'' core-collapse supernova model of \citet{fryer_2012}. In contrast to \citet{chawla_gaia_2022}, they adopt a common-envelope ejection efficiency of $\alpha = 5$, and draw BH natal kicks from a Maxwellian with $\sigma = 15$ km s$^{-1}$. As we discuss in Section~\ref{sec:roast}, these different choices in initial conditions and binary evolution parameters lead to a somewhat different BH+LC population from that predicted by \citet{chawla_gaia_2022}, even in the absence of dynamical effects.

\citet{di_carlo_young_2024} create synthetic MW BH-LC populations by seeding both isolated binaries (IBs) and their simulated SCs into the same m12i galaxy as \citet{chawla_gaia_2022}. To derive the metallicity-dependent star formation rate (SFR), they bin the star particles in metallicity and age. Then, they draw clusters with replacement in each bin until 10\% of the star formation in m12i is account for. The remaining 90\% is turned into massive IBs through direct conversion into star particles in each bin; for details, we direct the reader to \citet{di_carlo_young_2024}. To match the original cluster simulations, the IBs are given absolute metallicities $Z \in \{0.02, 0.002, 0.0002\}$ (i.e., representing the mass fraction of metals).

Each SC is simulated for 100 Myr with \texttt{NBODY6++GPU} and \texttt{MOBSE}. Then, both the isolated and SC binaries are evolved to present day with \texttt{COSMIC} \citep{breivik_cosmic_2020}, assuming the same parameters and recipes as in \citet{di_carlo_2020}. IBs are evolved from ZAMS to the current age of the universe (13.78 Gyr), while the evolution of SC binaries is restarted from the last recorded state in the $N$-body simulations and continued until present day. The final BH-LC population is restricted to systems with age $\geq 100$ Myr. \citet{di_carlo_young_2024} apply this cut for three reasons: first, their SCs are evolved for 100 Myr (so they cannot track systems formed in the past 100 Myr); second, the $z = 0$ SFR of the m12i galaxy is a few times higher than the observed MW SFR \citep[][]{hafen_2022}, meaning that the model is likely to overestimate the number of young binaries; and third, their work focuses on understanding the formation channels of Gaia BH1 and Gaia BH2, which are both Gyrs old. 

For the purpose of comparing formation channels, we only consider the population of BH-LC systems formed in the SCs of \citet{di_carlo_young_2024} when using the forward model of \citet{2024OJAp....7E.100E} to make predictions for detectable Gaia BHs. This includes both primordial binaries and dynamically exchanged binaries in those SCs (see Section \ref{sec:roast}). Nevertheless, to gain insight into the effect of differing input assumptions, we compare the IBE population of \citet{di_carlo_young_2024} against that of \citet{chawla_gaia_2022} in Section~\ref{sec:roast}. We note that removing systems with ages $< 100$ Myr from the IBE population of \citet{chawla_gaia_2022} reduces its size by $\sim 2\%$, but has a larger effect on the detectable population. We discuss the effects of this cut further in Section~\ref{sec:isolated}.

\subsection{Summary of Differences in Initial Conditions and Binary Evolution Modeling}

We summarize the key differences between the population synthesis models of \citet{chawla_gaia_2022} and \citet{di_carlo_young_2024} with respect to initial conditions and binary evolution modeling below.

\citet{chawla_gaia_2022} draw primary masses from a \citet{kroupa_2001} IMF. They adopt a uniform mass ratio distribution, with lower and upper limits of $0.08\,M_{\odot}$ and the primary mass on the secondary mass. Orbital eccentricities are drawn from a thermal distribution, and orbital semi-major axes are drawn from a log-uniform distribution with an upper limit of $10^{5}\,R_{\odot}$. The assumed initial binary fraction is $0.5$.

\citet{di_carlo_young_2024} also assume a \citet{kroupa_2001} IMF. Orbital periods and eccentricities are drawn according to the distributions of \citet{sana_binary_2012}, which leads to an upper limit on the initial period of $10^{3.5}$ d. Mass ratios are assigned differently for each environment. Secondary masses for binaries in SCs are assigned according to the mass ratio distribution of \citet{sana_binary_2012}, with a lower limit of $q = 0.1$. On the other hand, in the IBE population, secondary masses are drawn uniformly between $0.1\,M_{\odot}$ and the primary mass. The assumed initial binary fraction is $0.4$ in both the IBE and SC populations.

\citet{chawla_gaia_2022} adopt a common-envelope ejection efficiency of $\alpha = 1$, while \citet{di_carlo_young_2024} use $\alpha = 5$ instead. \citet{chawla_gaia_2022} assume that BH natal kicks are reduced by fallback relative to a 1D Maxwellian with $\sigma = 265$ km s$^{-1}$, while \citet{di_carlo_young_2024} draw BH natal kicks directly from a 1D Maxwellian with $\sigma = 15$ km s$^{-1}$. Unlike \citet{chawla_gaia_2022}, \citet{di_carlo_young_2024} restrict their final BH-LC population to systems with age $\geq 100$ Myr. We further explore how the differences in these assumptions impact the resulting isolated binary populations in Section \ref{sec:roast}. 

\subsection{Comparison of Intrinsic Populations}

\begin{figure*}
\epsscale{1.1}
\plotone{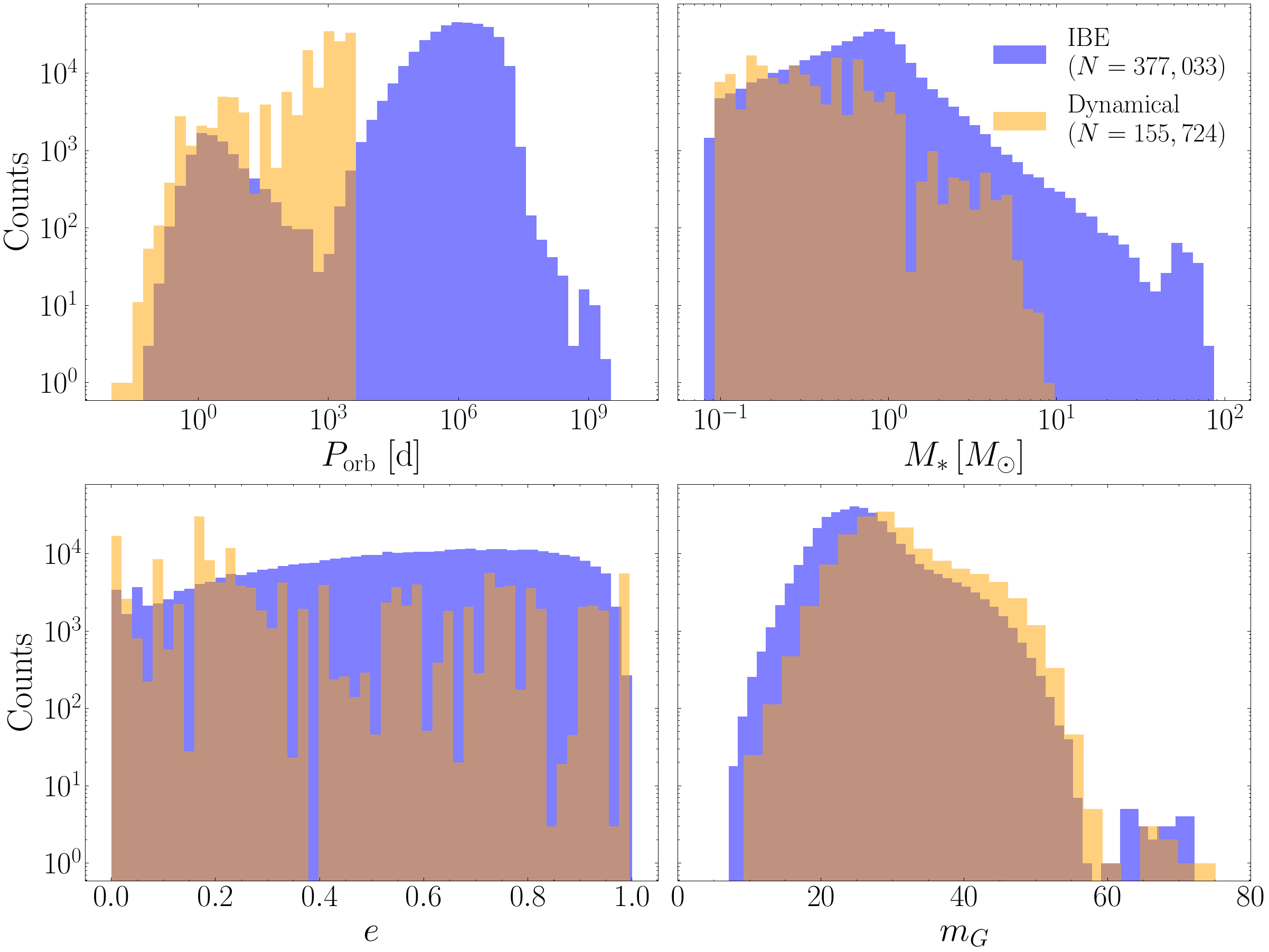}
\caption{Comparison of intrinsic physical properties of the isolated binary BH-LC population of \citet{chawla_gaia_2022} (blue) and the dynamically assembled BH-LC population of \citet{di_carlo_young_2024} (orange). The population of binaries formed via IBE features more LCs that are brighter and more massive relative to the population of dynamically assembled binaries. The orbital period distribution of the binaries formed via IBE is bimodal, while the orbital period distribution of the dynamically assembled binaries falls off steeply at $\sim10^{3.5}$ days.}
\label{fig:intrinsic_props}
\end{figure*}

We compare the apparent $G$-band magnitudes, orbital periods, luminous companion masses, and eccentricities of the isolated binary BH-LC population of \citet{chawla_gaia_2022} and the dynamically assembled BH-LC population of \citet{di_carlo_young_2024} in Figure~\ref{fig:intrinsic_props}. Here we show the full populations, without yet accounting for selection effects. The population of BH-LC binaries formed via IBE features systems with brighter apparent magnitudes and higher luminous companion masses than the dynamically assembled population. In addition, the period distribution of the population formed via IBE is bimodal, while the period distribution of the dynamical population falls off steeply at $\sim10^{3.5}$ days. The bimodal period distribution in the population formed via IBE arises because the population includes both short-period systems that went through common envelope evolution and long-period systems that never interacted. The drop-off above $\sim10^{3.5}$ days in the period distribution of the dynamical population is primarily a result of the adopted initial period distribution (see Section~\ref{sec:roast}). The period distribution of the dynamical population peaks in the range that \textit{Gaia} is most sensitive to, while the corresponding distribution for the population formed via IBE has a gap at those periods.

\begin{figure*}
\epsscale{1.15}
\plotone{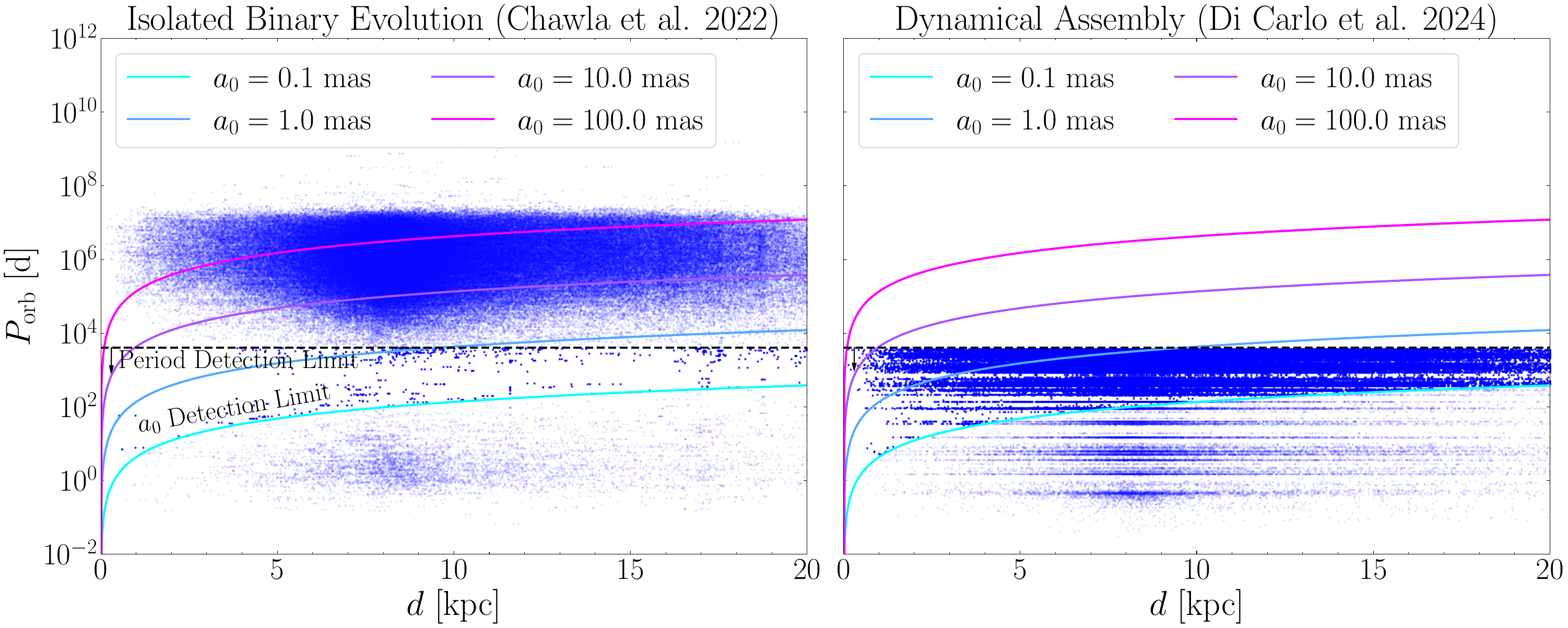}
\caption{Orbital period versus distance for Galactic populations of BH-LC systems formed via IBE (\citealt{chawla_gaia_2022}, left) and dynamical assembly (\citealt{di_carlo_young_2024}, right). Curves of constant angular semi-major axis of the photocenter $a_0$, assuming a typical BH mass of $9\,M_{\odot}$ and a typical LC mass of $1\,M_{\odot}$, are plotted as well. We show both the upper detectable limit on orbital period ($\sim 4000$ days) and the lower detectable limit on $a_0$ ($0.1$ mas) in DR4. Binaries that fall within these detectability limits are plotted with lower transparency. The IBE channel predicts a period gap at $\sim10^2$--$10^4$ days, which is precisely the period range to which \textit{Gaia} is most sensitive. In contrast, the dynamical assembly channel predicts many systems in this period range.}
\label{fig:both_sma}
\end{figure*}

We show scatter plots of orbital period $P_{\text{orb}}$ versus distance $d$ from the Sun for the isolated binary BH-LC population of \citet{chawla_gaia_2022} and the dynamically assembled BH-LC binary population of \citet{di_carlo_young_2024} in the left and right panels of Figure~\ref{fig:both_sma}, respectively. We overplot curves of constant angular semi-major axis of the photocenter $a_0$, assuming a typical BH mass of $9\,M_{\odot}$ and a typical LC mass of $1\,M_{\odot}$.  We show the DR4 detectability limits of $P_{\text{orb}} \lesssim 4000$ days (representing twice the observing baseline) and $a_0 > 0.1$ mas, with binaries that fall within these detectability limits plotted with lower transparency. Since Gaia BHs that are likely detectable in DR4 lie in the period gap in-between wide, non-interacting binaries ($P_{\text{orb}} \gtrsim 10^4$ days) and close, post-common envelope binaries ($P_{\text{orb}} \lesssim 10^2$ days), the dynamical channel produces more BH binaries in the period range to which \textit{Gaia} is most sensitive.

\subsection{Mock Observations}

For both formation channels, we bootstrap 99 additional populations by choosing different random locations of the solar neighborhood in the simulated m12i galaxy at a radius of 8.1 kpc, resulting in 100 total realizations. Then, we calculate extinctions $A_G$ and apparent magnitudes $m_G$ in the \textit{Gaia} G-band using the \texttt{Combined19} 3D dust map in the \texttt{mwdust} package \citep{bovy_2016}, which combines the maps of \citet{drimmel_2003}, \citet{marshall_2006}, and \citet{green_2019}. In doing so, we assume that $A_G = 2.8 E(B - V)$, where $E(B - V)$ is the color excess on the scale of \citet{schlegel_1998}. We assign a random orientation to each binary in each realization by sampling the cosine of the inclination, the argument of periastron, the longitude of the ascending node, and the periastron time from uniform distributions. To speed up our computations, we adopt preliminary cuts of $d < 33$ kpc, $a_0 > 0.1$ mas (for context, the tightest orbits published in DR3 have $a_0 \approx 0.2$ mas), $m_G < 19$ (sources with $m_G > 19$ were not considered for orbital solutions in DR3), and $P_{\text{orb}} < \{4000, 7000, 10000\}$ days in DR3, DR4, and DR5, respectively (since binaries with $P_{\text{orb}}$ significantly longer than the observing baseline are undetectable).\footnote{We verified that relaxing these preliminary cuts causes very few additional BH-LC systems to receive astrometric binary solutions, leading to no noticeable changes in our results.} 

Finally, we pass the synthesized populations through the \texttt{gaiamock}\footnote{\texttt{https://github.com/kareemelbadry/gaiamock}} pipeline of \citet{2024OJAp....7E.100E} to determine whether or not each BH-LC system receives an astrometric binary solution in each \textit{Gaia} data release. To construct a mock astrometric binary catalog, the \texttt{gaiamock} pipeline accepts a binary with specified parameters, generates mock observations using the \textit{Gaia} scanning law, and fits the epoch astrometry using \textit{Gaia}'s astrometric model cascade to identify what kind of solution (i.e., single-star, acceleration, or orbital) the binary will receive. We identify systems in the generated astrometric binary catalog as ``detected'' if they pass the detectability cuts in \textit{Gaia} DR3 (see \citealt{halbwachs_gaia_2023} for details). For DR4 and DR5, we also consider looser cuts of $\varpi / \sigma_{\varpi} > 5$ and $a_0 / \sigma_{a_0} > 5$, where $\varpi = 1 / d$ is the parallax and $\sigma_{\varpi}$ and $\sigma_{a_0}$ are the errors on the parallax and angular semi-major axis of the photocenter, respectively. We note that applying these looser cuts to real data is likely to result in a significant number of false-positives; however, these will be identifiable with spectroscopic follow-up. 

\begin{figure*}
\epsscale{1.15}
\plotone{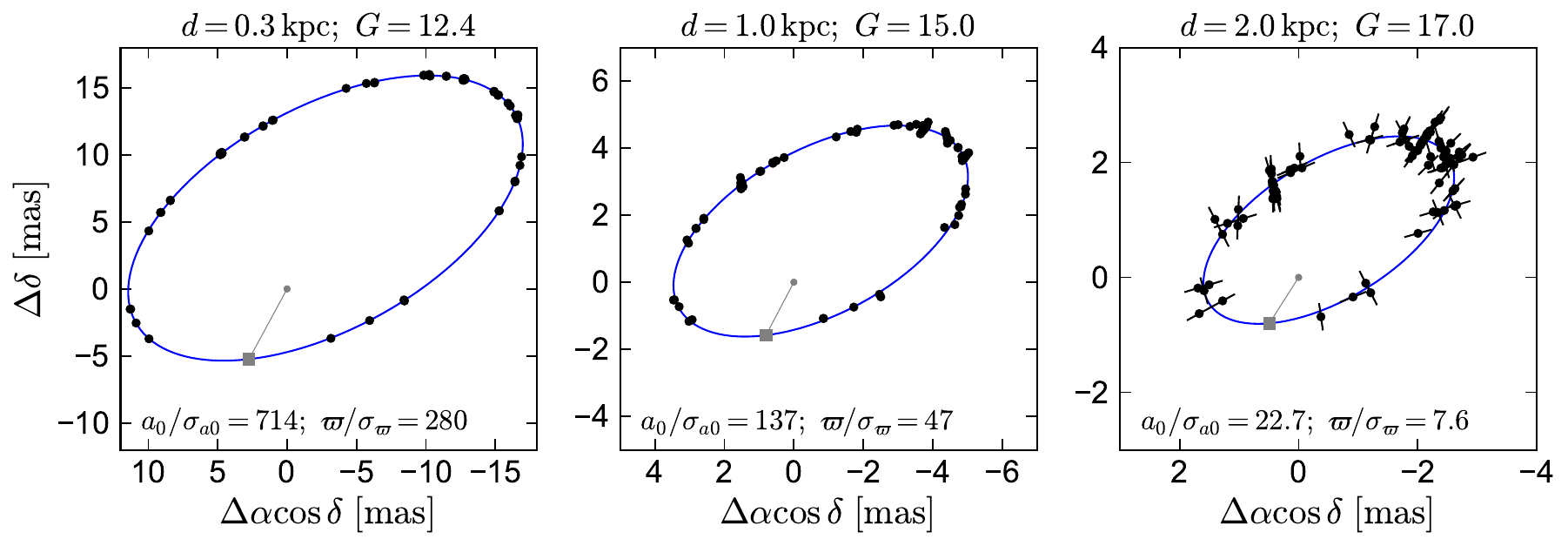}
\caption{DR4 observations of a typical BH-LC system featuring a solar-type star and a $10\,M_{\odot}$ BH in an orbit with $P_{\text{orb}} = 1500$ days, $e = 0.5$, and $i = 60^{\circ}$. Mock observations are generated at three different distances, with the orbital significance and uncertainty-normalized parallax both decreasing as the distance increases. In all cases, the \texttt{gaiamock} pipeline predicts that the binary will receive an orbital solution in DR4.}
\label{fig:ex_astro_fits}
\end{figure*}

We show DR4 observations of a typical BH-LC system featuring a solar-type star and a $10\,M_{\odot}$ BH in an orbit with $P_{\text{orb}} = 1500$ days, $e = 0.5$, and $i = 60^{\circ}$ in Figure~\ref{fig:ex_astro_fits}. We generate mock observations at three different distances. We find that the orbital significance and uncertainty-normalized parallax both decrease as the distance increases. At all distances shown, \texttt{gaiamock}  predicts that the binary will receive an orbital solution in DR4, with the binary at $d = 2.0$ kpc being close to the detectability limit of $\varpi / \sigma_{\varpi} > 5$. In all cases where the semi-major axis is $\gtrsim 1$ au, we expect $a_0 / \sigma_{a_0}$ to be greater than $\varpi / \sigma_{\varpi}$, so the latter quantity is more relevant to determining whether a given Gaia BH is predicted to receive an orbital solution in DR4.

\section{Results} \label{sec:results}

For each MW realization, we first select BH-LC systems that received astrometric binary solutions via the \texttt{gaiamock} pipeline. We identify ``typical'' MW realizations as those hosting the median number of detected BH-LC systems across 100 trials. 

\subsection{Comparison to Observations in DR3}
\label{sec:dr3}

\begin{figure*}
\epsscale{1.0}
\plotone{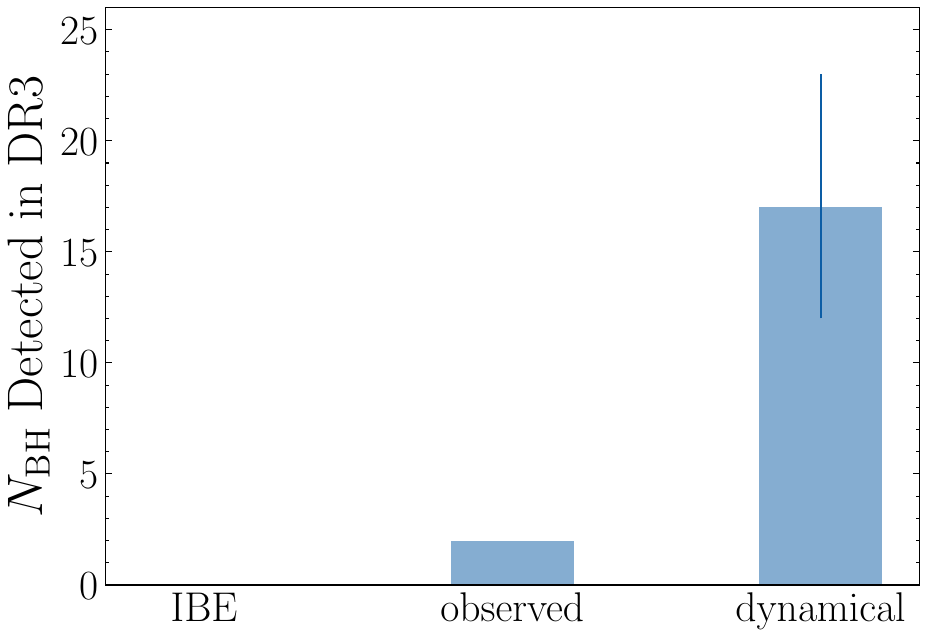}
\caption{Median number of detected BH-LC systems in \textit{Gaia} DR3 across 100 realizations. The error bar signifies the middle 68\% of counts. We compare predictions against observations and between both formation channels. Based on the observed number of BH-LC systems in DR3, we find that the IBE formation channel underestimates the number of detectable Gaia BHs. On the other hand, the dynamical formation channel overestimates the number of detectable Gaia BHs by a factor of $\sim 8$.}
\label{fig:dr3_bar_plot}
\end{figure*}

We show the median number across 100 realizations of Gaia BHs predicted to receive astrometric binary solutions in \textit{Gaia} DR3 in Figure~\ref{fig:dr3_bar_plot}. The lower and upper error bars represent the scatter across realizations, and are computed from the 16th and 84th percentiles of counts, respectively. We compare predictions between the two formation channels, and show the number of observed Gaia BHs ($N = 2$, since Gaia BH3 was discovered with DR4 data) as well. We find that the IBE formation channel investigated in \citet{chawla_gaia_2022} clearly underestimates the number of detectable Gaia BHs, as it predicts that no Gaia BHs are detectable in DR3 in the overwhelming majority of realizations. In fact, at least one BH-LC binary is detectable in only 8 out of 100 realizations. In these cases, the detectable LCs (which range from solar mass to highly massive) are all significantly more luminous than the Sun.

On the other hand, the dynamical formation channel investigated in \citet{di_carlo_young_2024} likely overestimates the number of detectable Gaia BHs. The median number across 100 realizations of Gaia BHs detected in the dynamical population in DR3 is $17_{-5}^{+6}$. 
Comparing to the number of systems already known to be detectable in DR3 (Gaia BH1 and Gaia BH2), we conclude that the dynamical formation channel over-predicts the number of detectable Gaia BHs by a factor of $\sim 8$.

We consider it unlikely that there are any more Gaia BHs waiting to be discovered in DR3 astrometric orbital solutions, since all promising candidates have been followed up spectroscopically \citep{el-badry_sun-like_2023, el-badry_red_2023, el-badry_population_2024}. 

\subsection{Isolated Binary Evolution Formation Channel}
\label{sec:isolated}

\begin{figure*}
\epsscale{1.15}
\plotone{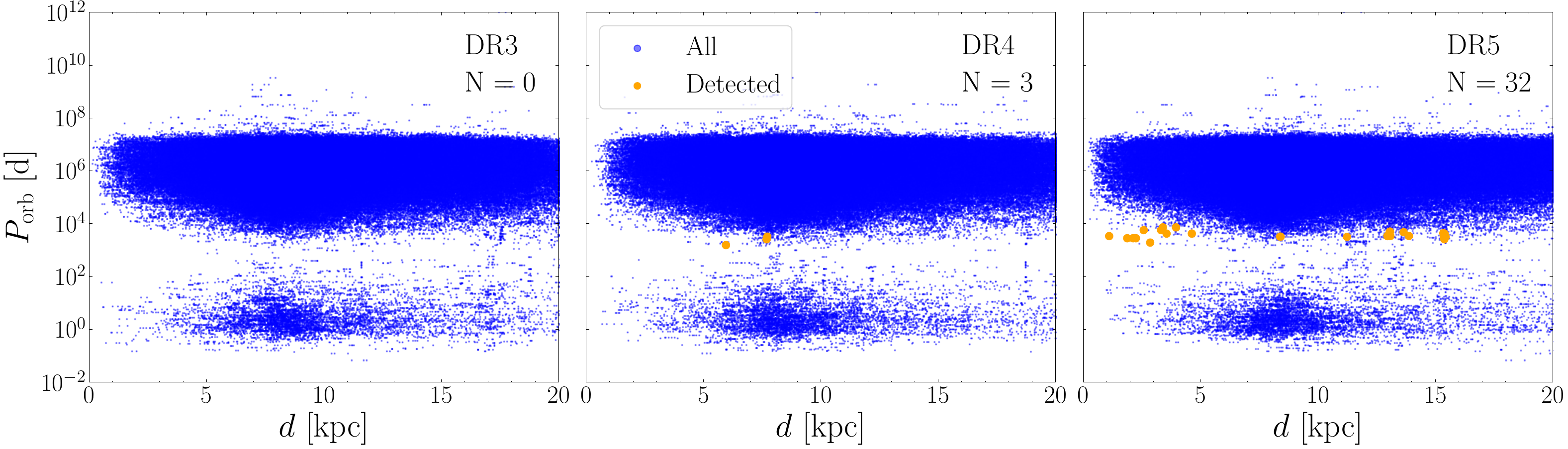}
\caption{Orbital period versus distance for typical realizations of the Milky Way BH-LC population formed via IBE in the model of \citet{chawla_gaia_2022}. BH-LC systems detected in \textit{Gaia} DR3, DR4, and DR5 are marked by orange circles in the left, middle, and right panels respectively. As expected, more systems are detectable at larger distances with a longer time baseline. {\it Gaia} is most sensitive to binaries with periods of $10^2$--$10^4$ days -- precisely where the model predicts a gap separating close binaries that have gone through a common envelope event and wide binaries that have not interacted.}
\label{fig:isolated_scatter}
\end{figure*}

We plot orbital period versus distance for the BH-LC binaries formed via IBE in typical realizations for DR3, DR4, and DR5 in Figure~\ref{fig:isolated_scatter}. We see that longer time baselines allow for more systems to be detected at larger distances. However, as mentioned above, all of the detected systems have orbital periods of $10^2$--$10^4$ days, which falls in the gap separating wide, non-interacting binaries and close, post-common envelope evolution binaries in the IBE population. The upper limit reflects \textit{Gaia}'s finite observing baseline, while the lower limit reflects a steep drop-off in the model's period distribution toward shorter periods. All detectable BH-LC binaries fall on the long-period edge of the gap, and are presumably systems that never interacted.

\begin{figure*}
\epsscale{1.0}
\plotone{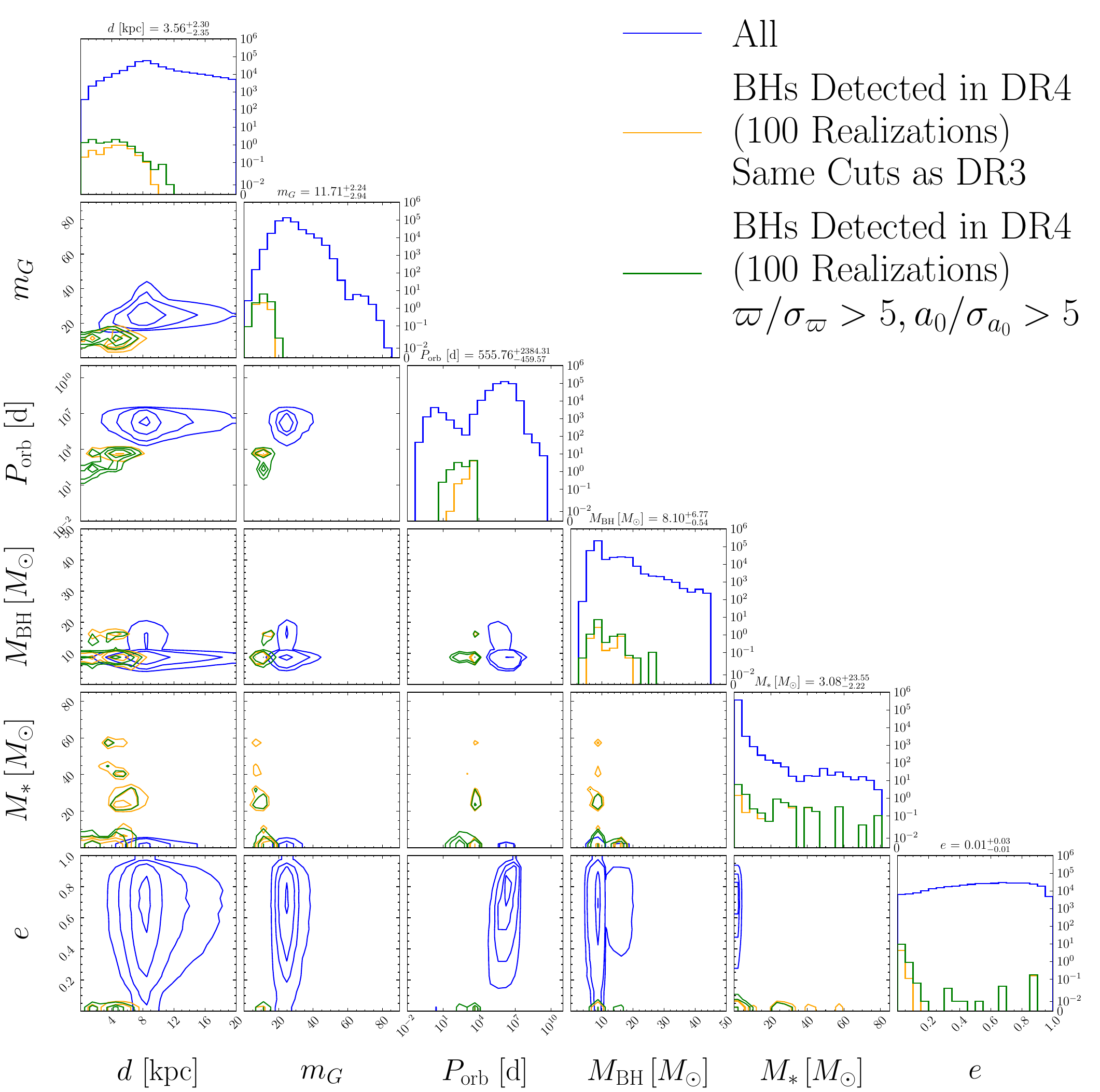}
\caption{Properties of BH-LC systems formed via IBE. The diagonal entries display the marginal distribution of each parameter, while each of the other panels displays a joint distribution. Contours identify regions of high density in each parameter space. To reduce shot noise, we combine predictions from 100 realizations, each corresponding to a different placement of the Sun in the simulated galaxy. We then rescale the marginal distributions by $1/100$, so that the counts show the average number of BHs per realization. The distributions for the entire population are plotted in blue, while the distributions for the subset detected in DR4 (assuming the same quality cuts as in DR3) are plotted in orange. We see that detected systems tend to have low orbital eccentricities, with an observational bias toward brighter and more massive stars. We plot the distributions for the subset of BH-LC systems detected in DR4 assuming less stringent quality cuts in green. We observe similar trends, though adopting looser cuts leads to more detectable Gaia BHs.}
\label{fig:isolated_corner}
\end{figure*}

We show a corner plot of physical parameters for the BH-LC systems formed via IBE in Figure~\ref{fig:isolated_corner}. To reduce sampling noise, we stack all 100 realizations, with the distribution for the entire population plotted in blue, and the distribution for the detected population of Gaia BHs (assuming the same cuts as DR3) in orange. We also show the distribution of Gaia BHs detected assuming simpler cuts of $\varpi / \sigma_{\varpi} > 5$ and $a_0 / \sigma_{a_0} > 5$ in green. We rescale all marginal distributions by $1/100$. The detected population of Gaia BHs is biased towards brighter and more massive luminous companions. We also find that less eccentric Gaia BHs are preferentially detected. Assuming simpler detectability cuts increases the number of detected BH-LC systems, but does not change these observed trends. 

If we restrict the population to BH-LC systems older than 100 Myr (for consistency with \citealt{di_carlo_young_2024}), the median number of BHs detected in DR4 drops from $3^{+6}_{-3}$ to $1^{+2}_{-1}$. This is because many of the detectable BH-LC systems in the population of \citet{chawla_gaia_2022} host massive, young luminous stars. If we require that Gaia BHs be older than 100 Myr, then the maximum LC mass decreases from $\approx70\,M_{\odot}$ to $\approx4\,M_{\odot}$.

\subsection{Dynamical Formation Channel}
\label{sec:dynamical}

\begin{figure*}
\epsscale{1.15}
\plotone{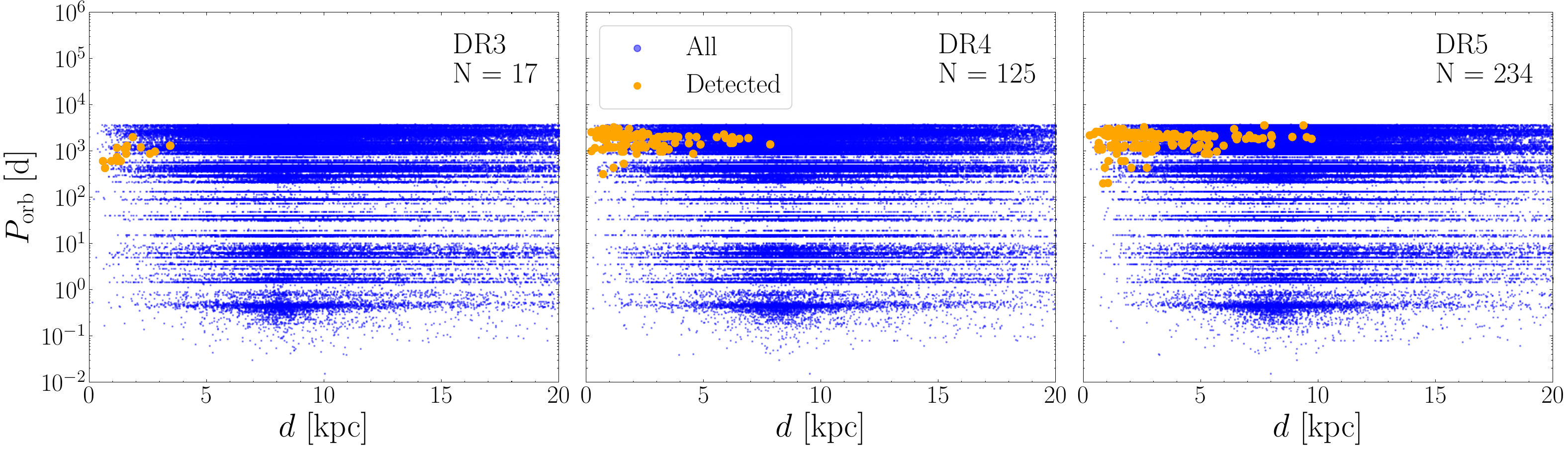}
\caption{Same as Figure \ref{fig:isolated_scatter}, but for typical realizations of the population formed via dynamical interactions in the model of \citet{di_carlo_young_2024}. Once again, more systems are detectable at larger distances with a longer time baseline. Almost all detected systems have orbital periods in the range of $10^2$--$10^4$ days. Far more systems are predicted to be detectable in the dynamical formation channel than in the isolated formation channel, due to many more systems originally existing in this range of orbital separations.}
\label{fig:dynamic_scatter}
\end{figure*}

We plot orbital period versus distance for the dynamically formed BH-LC systems in typical model realizations for DR3, DR4, and DR5 in Figure~\ref{fig:dynamic_scatter}. As before, we find that longer time baselines lead to more systems being detected at larger distances. However, in this case, there are far more systems in the range of  $P_{\text{orb}} = 10^2$--$10^4$ days, leading to many more Gaia BHs on average.

\begin{figure*}
\epsscale{1.0}
\plotone{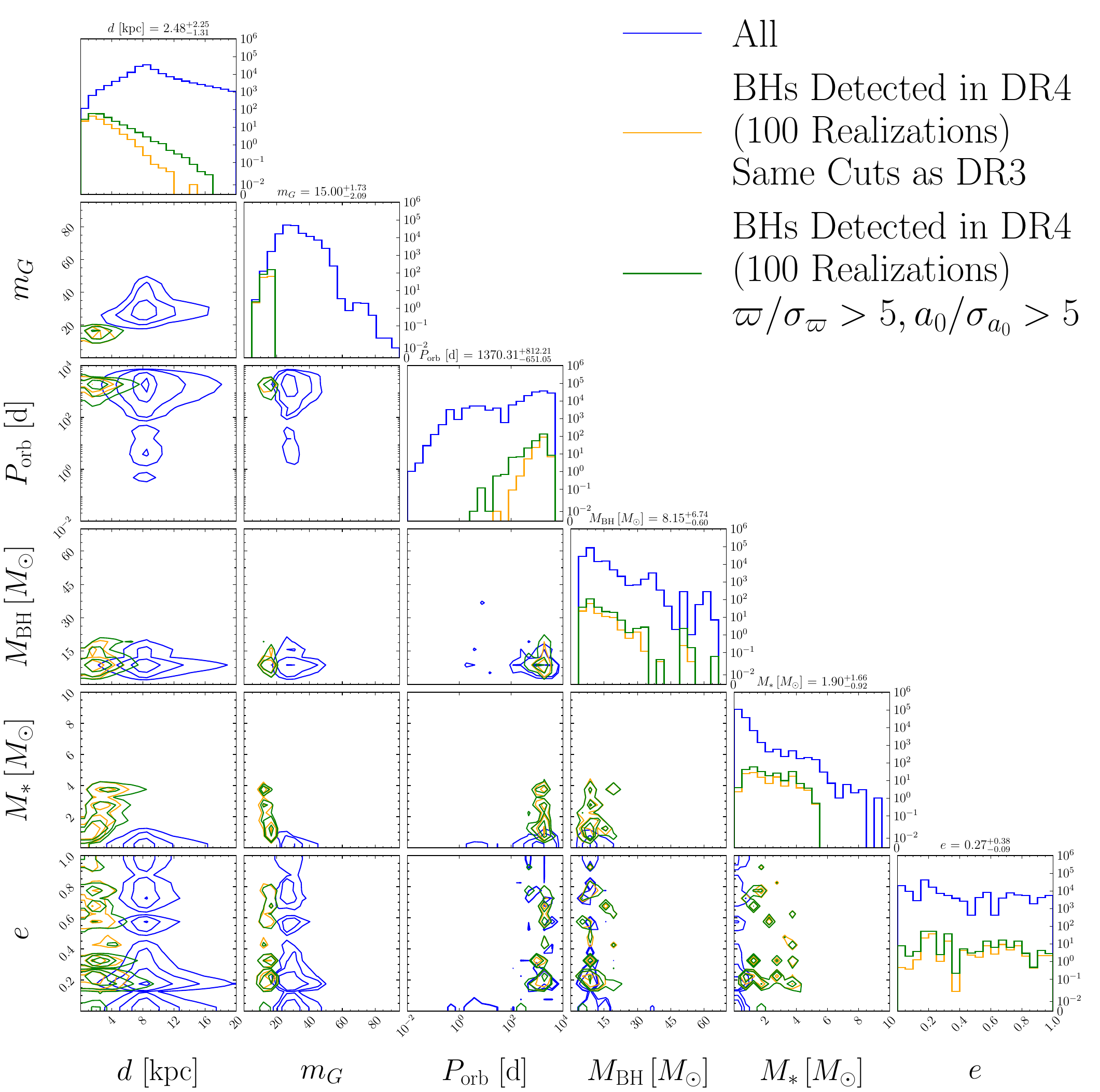}
\caption{Same as Figure \ref{fig:isolated_corner}, but for BH-LC systems formed via dynamical interactions. We once again find the detected systems to be biased towards brighter and more massive luminous companions. However, we now detect systems at a range of eccentricities. In the subset of BH-LC systems detected in DR4 assuming less stringent cuts, we observe similar trends, though adopting those cuts leads to more detectable Gaia BHs.}
\label{fig:dynamic_corner}
\end{figure*}

We show a corner plot of physical parameters for the BH-LC systems formed via dynamical assembly in Figure \ref{fig:dynamic_corner}. We stack all 100 realizations, with the colors representing the same populations as in Figure \ref{fig:isolated_corner}. We rescale all marginal distributions by $1/100$. We once again find that the detected population of Gaia BHs is biased toward brighter and more massive luminous companions. However, unlike in the IBE channel, the population of detected Gaia BHs has support at all eccentricities. These observed trends do not change if simpler detectability cuts are assumed. 

\subsection{Comparison of Galactic Distributions}

\begin{figure*}
\epsscale{1.0}
\plotone{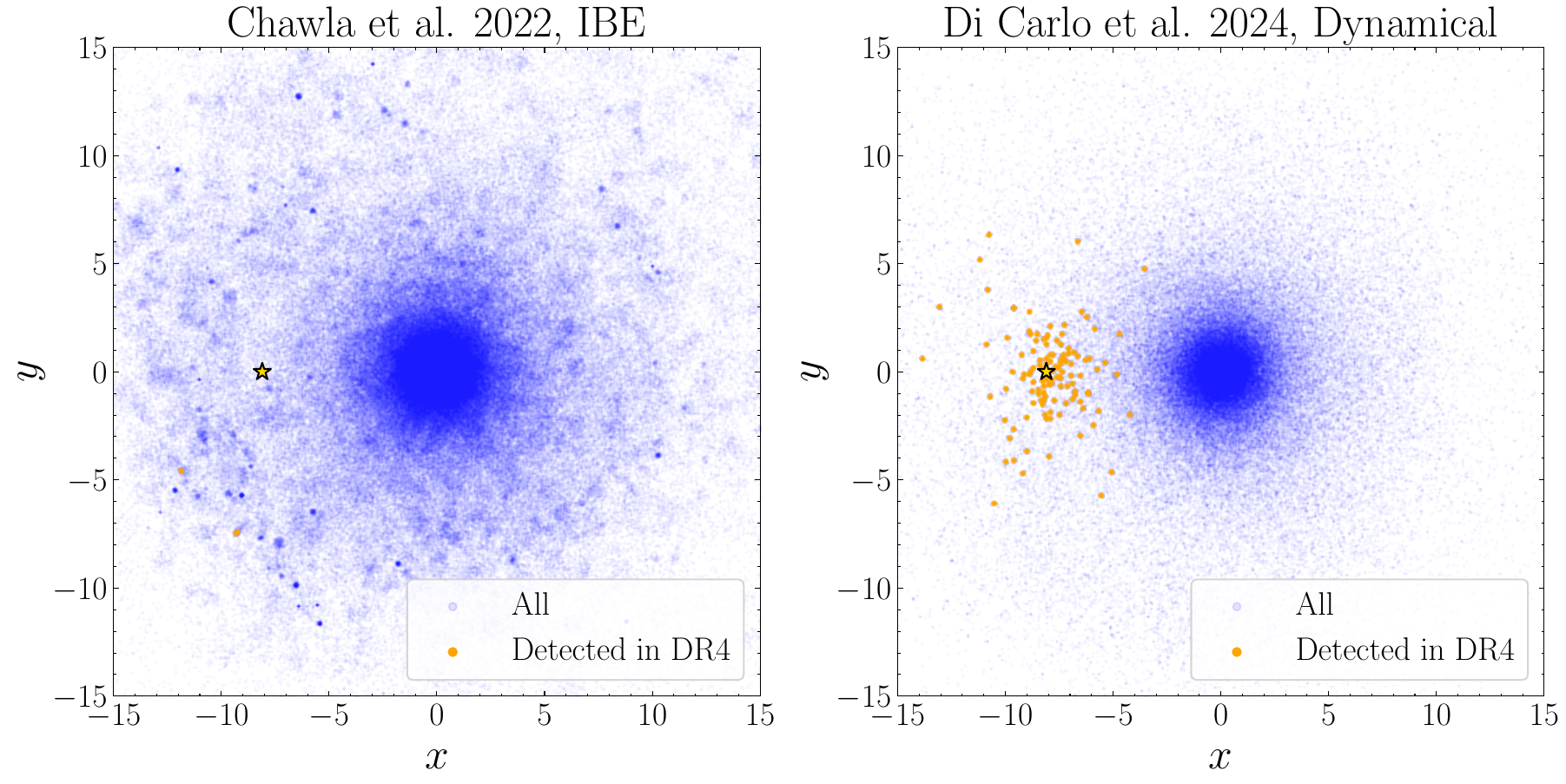}
\caption{Predicted spatial distribution of BH-LC populations formed via IBE (\citealt{chawla_gaia_2022}, left panel) and dynamical assembly (\citealt{di_carlo_young_2024}, right panel). We show all systems in blue, and the Gaia BHs detected in DR4 (for a typical MW realization) in orange. Because the IBE population from \citet{chawla_gaia_2022} includes BHs with young and massive companions, it is predicted to be clumpier than the dynamically assembled population from \citet{di_carlo_young_2024}. This causes more scatter in the number of Gaia BHs predicted to be detectable across model realizations. Meanwhile, the dynamically assembled BH-LC systems are smoothly distributed throughout the MW, tracing older stellar populations. The dynamically formed Gaia BHs that are detected in DR4 are clustered around the location of the Sun, which is marked with a gold star in both panels.}
\label{fig:mw_dist_plot}
\end{figure*}

We compare the Galactic distributions in Cartesian coordinates of the isolated binary BH-LC population of \citet{chawla_gaia_2022} and the dynamically assembled BH-LC population of \citet{di_carlo_young_2024} in Figure~\ref{fig:mw_dist_plot}. We plot the entire population of binaries in blue, and the systems that receive astrometric binary solutions in DR4 in a typical MW realization in orange. Because the isolated binary population from \citet{chawla_gaia_2022} includes BHs with young and massive LCs, its spatial distribution is predicted to be clumpier than that of the dynamical population of \citet{di_carlo_young_2024}, leading to more (non-Gaussian) scatter in the number of potentially detectable BH-LC systems across MW realizations. Meanwhile, the dynamically assembled BH-LC systems are smoothly distributed throughout the Galaxy, tracing older stellar populations, with potentially detectable BH-LC systems being clustered around the location of the Sun.

\subsection{Summary of Predictions}

\begin{figure*}
\epsscale{1.1}
\plotone{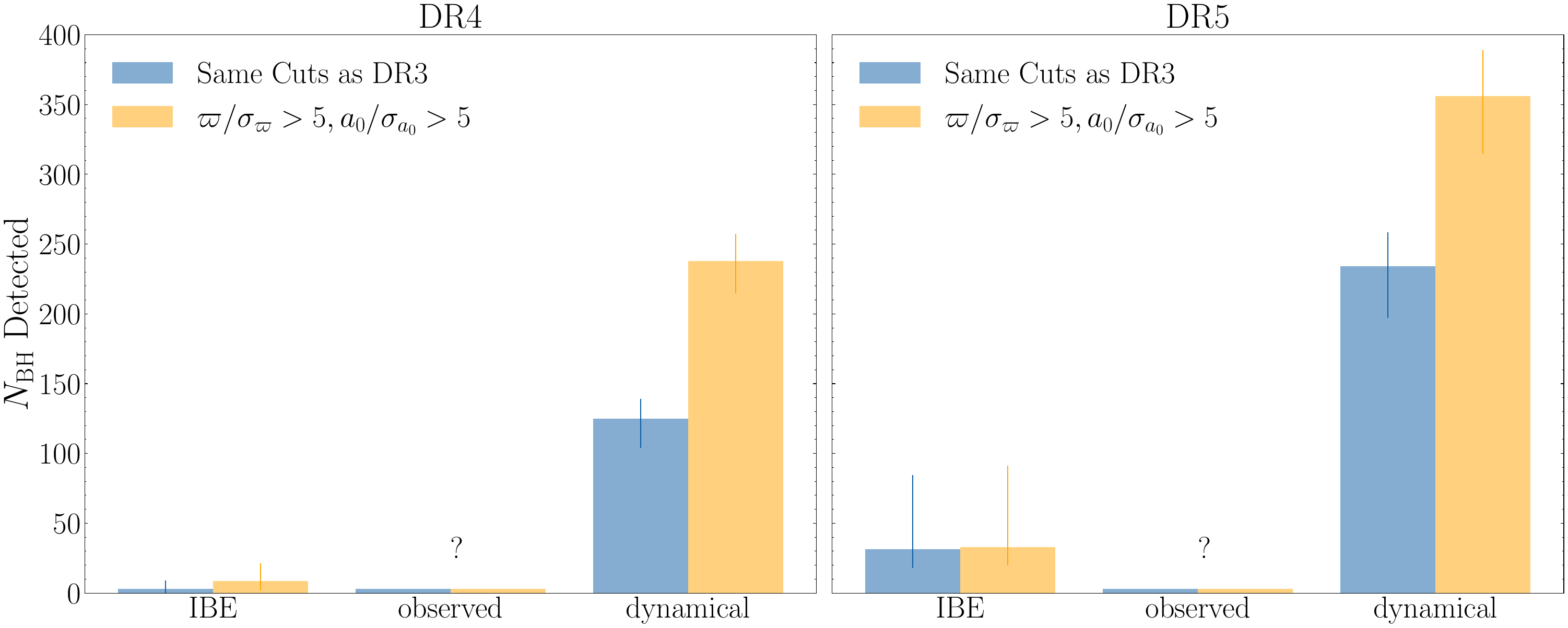}
\caption{Number of detected BH-LC systems in future \textit{Gaia} data releases. We report the median number of Gaia BHs detected across 100 realizations, with error bars derived from the 16th and 84th percentiles. We compare predictions against observations and between both formation channels. We plot the results assuming the same cuts as \textit{Gaia} DR3 in blue, and the results assuming less stringent detectability cuts in orange. Assuming these simpler cuts increases the number of detectable systems by a factor of $\sim2$. Based on their predictions for DR3, we assume that the isolated binary model from \citet{chawla_gaia_2022} underestimates the number of detectable BHs, while the dynamical model of \citet{di_carlo_young_2024} overestimates it by a factor of $\sim8$. On this basis, we predict that (using the simpler cuts) $30_{-3}^{+2}$ Gaia BHs will be discovered in DR4 and $45_{-5}^{+4}$ Gaia BHs will be discovered in Gaia DR5.}
\label{fig:bar_plot}
\end{figure*}

We summarize the number of Gaia BHs predicted to receive astrometric binary solutions in future \textit{Gaia} data releases in Figure~\ref{fig:bar_plot}. We report the median number of Gaia BHs detected across 100 realizations, with error bars derived from the 16th and 84th percentiles. We compare predictions between the two formation channels, and show the number of observed Gaia BHs as well. We plot the results assuming the same cuts as DR3 in blue, and the results assuming the simpler detectability cuts of $\varpi / \sigma_{\varpi} > 5$ and $a_0 / \sigma_{a_0} > 5$ in orange. Using the simpler cuts increases the number of detected Gaia BHs by a factor of $\sim 2$. 

We find that Gaia BHs tend to have orbital periods in the range of $10^2$--$10^4$ days, regardless of formation channel. The detected BHs in isolated binaries have masses of $3$--$25\,M_{\odot}$, while those in dynamically assembled binaries have masses of $5$--$50\,M_{\odot}$. 
In the dynamical population of \citet{di_carlo_young_2024}, the luminous companions to detected BHs have masses of $0.1$--$5\,M_{\odot}$, and the median apparent magnitude is $m_G \approx 15$. Meanwhile, in the isolated binary population of \citet{chawla_gaia_2022} (which does not require that systems be older than $100$ Myr), the luminous companions to detected BHs have masses up to $80\,M_{\odot}$, and the median apparent magnitude is $m_G \approx 12$. In both formation channels, the BH-LC systems are predicted to lie at typical distances of $1$--$5$ kpc from the Sun.

\subsection{Realistic Predictions for DR4 and DR5}

Since the IBE channel investigated by \citet{chawla_gaia_2022} significantly underestimates the number of detectable Gaia BHs, we adopt the dynamical assembly channel investigated by \citet{di_carlo_young_2024} as our fiducial formation pathway when making predictions for the number of Gaia BHs that will be discovered in future \textit{Gaia} data releases. Based on the analysis in Section \ref{sec:dr3}, we suggest that this dynamical channel overpredicts the number of detectable Gaia BHs by a factor of $\sim 8$, but produces a period distribution that is consistent with observations so far. If we divide the number of dynamically formed BH-LC systems that are detected in future data releases (see Figure~\ref{fig:bar_plot} for the raw numbers) by 8, we predict that (assuming the same detectability cuts as DR3) around $16_{-3}^{+2}$ Gaia BHs will be discovered in \textit{Gaia} DR4 and around $29_{-5}^{+3}$ Gaia BHs will be discovered in \textit{Gaia} DR5. If we assume simpler detectability cuts of $\varpi / \sigma_{\varpi} > 5$ and $a_0 / \sigma_{a_0} > 5$, the number of detectable Gaia BHs increases to $30_{-3}^{+2}$ in DR4 and $45_{-5}^{+4}$ in DR5, respectively. We note that our prediction for the number of detectable Gaia BHs in DR5 is probably an underestimate, because the sharp drop off at $\sim 10^{3.5}$ days in the period distribution of the dynamical population is likely unrealistic. This caveat only applies to DR5, because DR5 is the first data release in which we will be sensitive to such long periods.

\subsection{Searches for BHs with Acceleration Solutions}

An acceleration solution is the simplest extension of the single-star solution, and is appropriate when the photocenter motion is satisfactorily described by constant (i.e., for 7-parameter solutions) or variable (i.e., for 9-parameter solutions) acceleration terms in the right ascension and declination directions. This is expected for binaries with orbital periods that are at least a few times longer than the observing baseline. Acceleration solutions are placed before orbital solutions in the DR3 astrometric cascade, with 9-parameter solutions being accepted when:

\begin{eqnarray}
\label{eq:nine_param}
\cases{
    s_9 > 12 & \cr
    F_2 < 25 & \cr
    \varpi / \sigma_{\varpi} > 2.1 s_9^{1.05} & \cr
}
\end{eqnarray}

and 7-parameter solutions (i.e., for binaries that do not pass Equation~\ref{eq:nine_param}) being accepted when:

\begin{eqnarray}
\label{eq:seven_param}
\cases{
    s_7 > 12 & \cr
    F_2 < 25 & \cr
    \varpi / \sigma_{\varpi} > 1.2 s_7^{1.05} & \cr
}
\end{eqnarray}

where $\varpi / \sigma_{\varpi}$ is the uncertainty-normalized parallax, $s$ is the solution significance, and the $F_2$ statistic represents the goodness-of-fit \citep{2024OJAp....7E.100E}.

\begin{figure*}
\epsscale{1.15}
\plotone{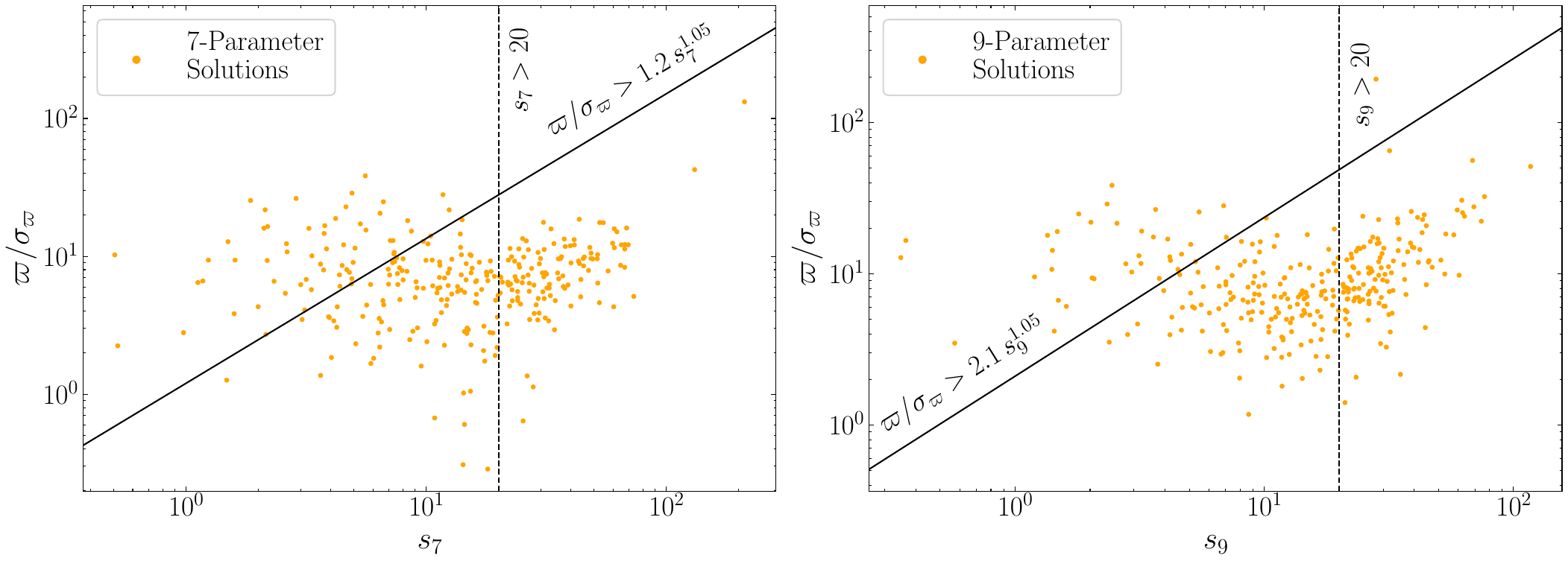}
\caption{Uncertainty-normalized parallax of the fitted 7-parameter and 9-parameter acceleration solutions as a function of significance of the astrometric solution for dynamically formed BH-LC systems with RUWE $> 1.4$ in DR4. We overplot the cuts on these quantities that were imposed in DR3. We see that no 7-parameter solutions and only one 9-parameter solution would be published in DR4 if the same cuts are assumed, implying that acceleration solutions are not a promising avenue for searches for dormant BHs, at least when the DR3 significance cuts are employed.}
\label{fig:acc_plot}
\end{figure*}

We plot $\varpi / \sigma_{\varpi}$ as a function of acceleration solution significance for dynamically formed BH-LC systems with RUWE $> 1.4$ in DR4 in Figure~\ref{fig:acc_plot}. We investigate both 7-parameter and 9-parameter fits, and show the cuts used for acceleration solutions in DR3. We see that, under these cuts, no 7-parameter solutions and only one 9-parameter solution would be published in DR4. This suggests that, at least with the cuts implemented in DR3, acceleration solutions are not a promising method to find dormant BHs.  We can explain this as follows: since BHs are more massive than Sun-like stars, BH-LC systems tend to lie at larger distances and lower parallaxes than LC-LC systems at the same acceleration significance, causing them to fall beneath the cut on uncertainty-normalized parallax used in \textit{Gaia} DR3.

\subsection{Searches for BHs with RUWE}

\begin{figure*}
\epsscale{1.0}
\plotone{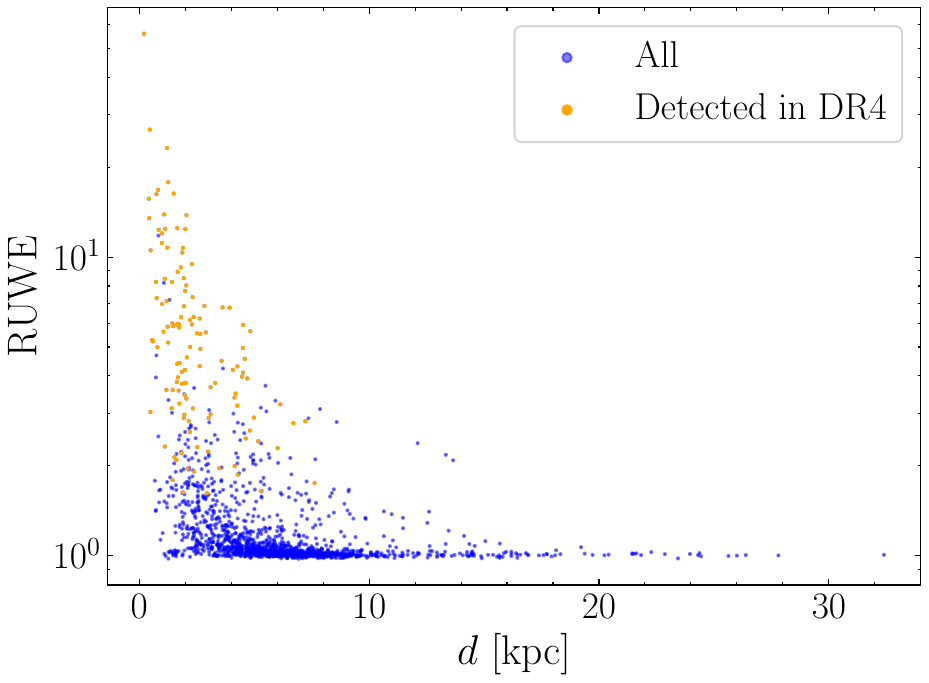}
\caption{RUWE as a function of distance for dynamically formed BH-LC systems in DR4. We denote the systems that receive an astrometric binary solution in orange. We see that the detected systems tend to have small distances and large RUWE values, with most binaries beyond $\sim5$ kpc predicted to have RUWE $< 1.4$. Some systems with large RUWE do not receive an astrometric binary solution because their orbital significance is too low or their fractional parallax uncertainty is too high, causing them to fail the quality cuts used to eliminate spurious solutions in DR3.}
\label{fig:ruwe_plot}
\end{figure*}

We plot RUWE in DR4 as a function of distance for dynamically assembled BH-LC systems in Figure~\ref{fig:ruwe_plot}. We plot BH-LC systems that receive an astrometric binary solution in orange. These detected systems tend to have large RUWE and lie at small distances. There are many false negatives at RUWE of $2$--$5$, implying that RUWE searches for Gaia BHs can find systems that did not receive an astrometric binary solution in DR4. About one-third of false negatives did not receive a solution because their orbital significance (i.e., the ratio of the photocenter semi-major axis to its uncertainty) was $< 5$. Almost all of the remainder of the false negatives did not receive a solution because their fractional parallax uncertainty failed the following period-dependent cut:

\begin{equation}
    \frac{\varpi}{\sigma_{\varpi}} > \frac{20000 \text{ d}}{P_{\text{orb}}}
\end{equation}

used by \citet{halbwachs_gaia_2023} to eliminate spurious solutions in \textit{Gaia} DR3. However, despite failing the period-dependent cut, the vast majority of these systems would actually receive an astrometric solution if the simpler cut of $\varpi/\sigma_{\varpi} > 5$ were used instead. We advocate using the simpler cut, since false positives can be eliminated via spectroscopic follow-up.

\section{Discussion} \label{sec:discussion}

\subsection{Caveats}

Our calibrated predictions for the number of Gaia BHs detectable in future \textit{Gaia} data releases depends on the assumption that most BH-LC systems in the Milky Way are dynamically assembled in open clusters. Given the significant uncertainties in the input physics to population synthesis models, including the efficiency of envelope ejection, the remnant mass prescription, and the strength of BH natal kicks, this assumption remains to be verified. As discussed in Section~\ref{sec:intro}, many alternative theoretical models for Gaia BH formation involving IBE or hierarchical triple dynamics have been proposed in the recent literature, and current observational evidence is insufficient to distinguish between these formation channels. Furthermore, based on the discoveries of Gaia BH1 and Gaia BH2 in DR3, we find that the dynamical formation channel proposed by \citet{di_carlo_young_2024} overestimates the number of detectable Gaia BHs by a factor of $\sim 8$; this overestimation factor should only be expected to remain constant in future data releases if the period distribution in the model population is realistic.

The \texttt{gaiamock} pipeline used in this work aims to forward-model the catalog of orbital solutions in DR3 from the \textit{Gaia} scanning law. As such, the astrometric cascade that is implemented is somewhat simplified compared to the one that \textit{Gaia} actually uses; nevertheless, the mock catalog provided in \citet{2024OJAp....7E.100E} is quite similar to the actual DR3 catalog, with the only the simulated eccentricity distribution being in tension with observations. In our work, we have investigated applying both the published \textit{Gaia} DR3 cuts and simplified detectability cuts in our post-processing; however, different quality cuts may be implemented in future releases. Since epoch astrometry will be published for all sources, the community will be able to construct independent astrometric orbit catalogs from DR4 data and apply less stringent cuts, such as those suggested in Figure~\ref{fig:bar_plot}. Furthermore, the per-epoch astrometric uncertainties in DR4 and DR5 may be smaller than in DR3, particularly at $G \lesssim 13$, where they are currently limited by systematics rather than photon noise. If treatment of systematics improves, the number of detectable BHs will increase; thus, our predictions are relatively conservative.

\subsection{What Fraction of MW Stars have BH Companions in au-scale Orbits?}

The dynamically assembled MW BH-LC population of \citet{di_carlo_young_2024} contains $155,724$ BH-LC systems. To calibrate to DR3, we divide this number by 8, reducing the size of the population to around $19,466$ binaries. Hence, if we estimate (to order of magnitude) that the MW contains $\sim100$ billion stars, then about one in $10$ million stars in the MW should orbit a BH in an au-scale orbit, regardless of whether the BH-LC binary is detectable with \textit{Gaia} astrometry or not.

Depending on whether we adopt simple signal-to-noise detectability cuts or the more stringent detectability cuts in DR3, we predict that $30^{+2}_{-3}$ Gaia BHs or $16^{+2}_{-3}$ Gaia BHs will be discovered in DR4, respectively. This indicates that we will be able to use astrometric orbital solutions in DR4 to detect about 1 in 1000 MW BHs with luminous companions in au-scale orbits. 

\subsection{Importance of the Adopted Binary Evolution Model }
\label{sec:roast}

\begin{figure*}
\epsscale{1.1}
\plotone{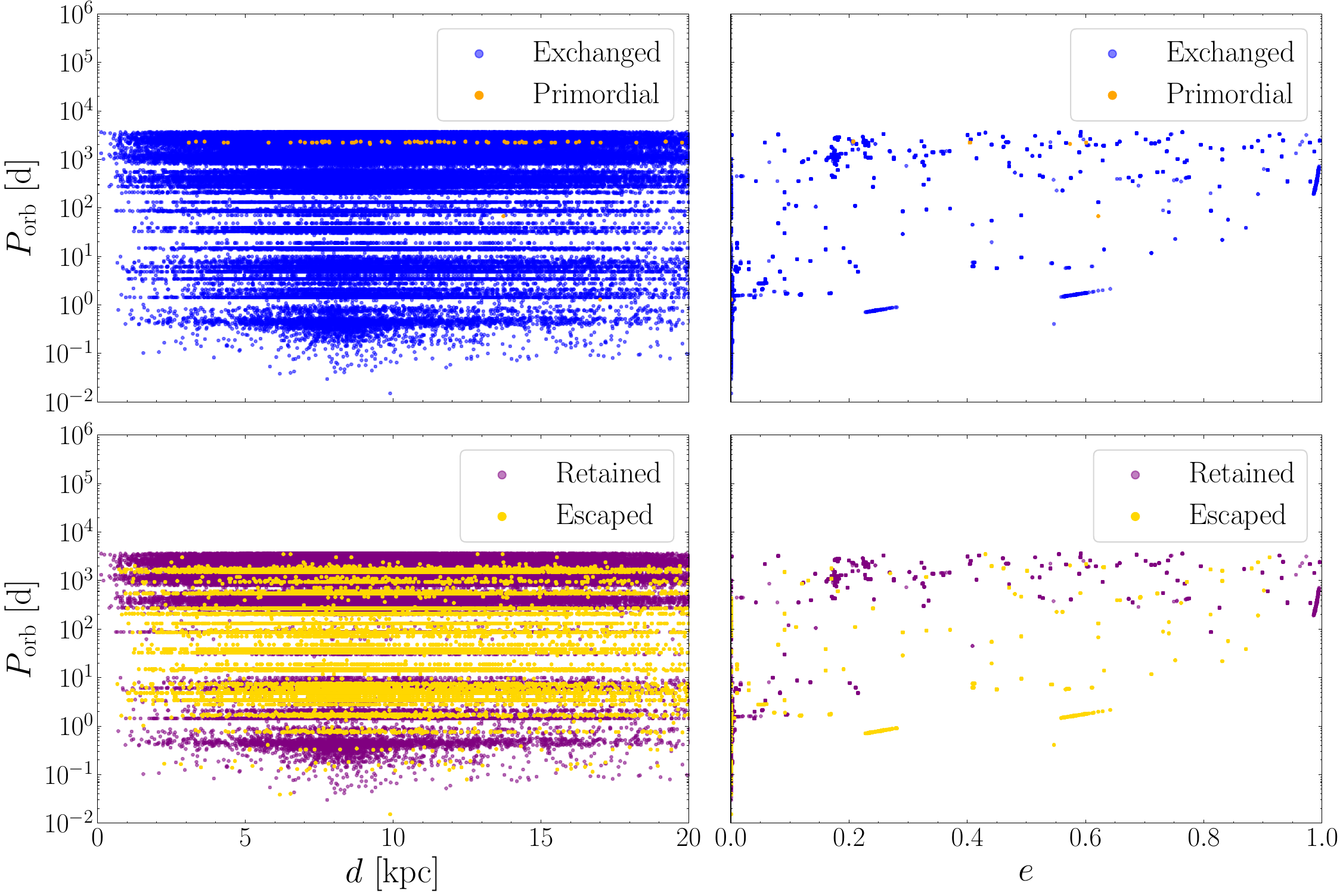}
\caption{Top Row: Orbital period versus distance (left) and eccentricity (right) for primordial (orange) and exchanged (blue) binaries in the population of BH-LC systems formed in the SCs of \citet{di_carlo_young_2024}. Bottom Row: Same as the top row, but separated based on whether the BH-LC binaries escape from (gold) or are retained in (purple) the open clusters in which they form after 100 Myr. We observe that, under the assumptions of \citet{di_carlo_young_2024}, most BH-LC systems form via dynamical exchanges and are retained in their open clusters for at least 100 Myr.}
\label{fig:dyn_prop_plot}
\end{figure*}

\begin{figure*}
\epsscale{1.1}
\plotone{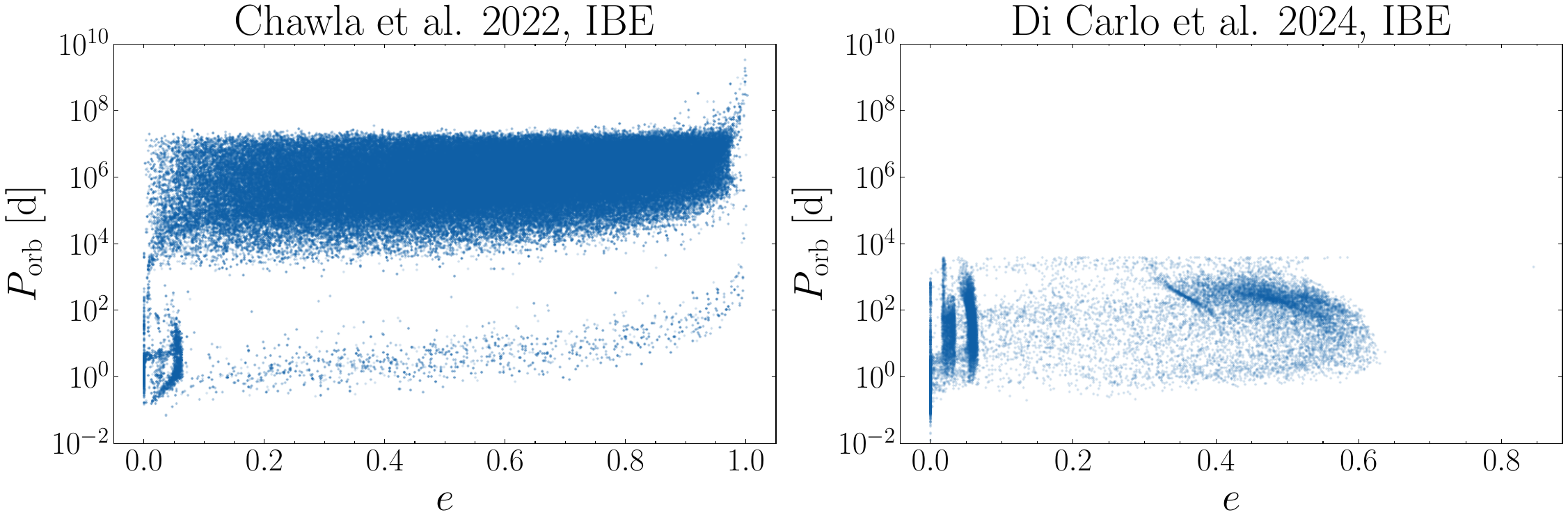}
\caption{Orbital period versus eccentricity for the populations formed via IBE from \citet{chawla_gaia_2022} (left panel) and \citet{di_carlo_young_2024} (right panel). The IBE population predicted by \citet{di_carlo_young_2024} lacks long-period binaries that did not interact, due to its initial period upper limit of $10^{3.5}$ d. Owing to its weaker kicks and larger common envelope $\alpha$, it contains post-common envelope binaries with longer periods and lower eccentricities than the \citet{chawla_gaia_2022} population. Most of the post-interaction binaries predicted by both IBE models have orbital periods too short to be detectable with {\it Gaia} astrometry.
}
\label{fig:compare_pop_plot}
\end{figure*}

As explained in Section \ref{sec:dicarlo}, \citet{di_carlo_young_2024} synthesize a population of BH-LC binaries formed via IBE in addition to a population of BH-LC binaries that are dynamically assembled in clusters. Applying the \texttt{gaiamock} pipeline to this population and assuming the same detectability cuts as in DR3, we find that $18^{+4}_{-3}$ BH-LC systems are predicted to be detectable in DR4 in the IBE population of \citet{di_carlo_young_2024}, while only $3^{+6}_{-3}$ BH-LC systems are predicted to be detectable in DR4 in the IBE population of \citet{chawla_gaia_2022}. While these results support the claim of \citet{di_carlo_young_2024} that young SCs dominate the production of BH-LC binaries, they also suggest that the physical characteristics of the two IBE populations are  different. In this section, we compare the IBE populations of \citet{di_carlo_young_2024} and \citet{chawla_gaia_2022} to assess how much the differences in their population synthesis assumptions matter.

We plot orbital period versus distance and eccentricity for the BH-LC systems formed in the SCs of \citet{di_carlo_young_2024} in Figure~\ref{fig:dyn_prop_plot}. In the top row, we plot primordial binaries in orange and dynamically exchanged binaries in blue. We find that most Gaia BHs are formed via dynamical exchanges. In the bottom row, we plot binaries that escaped from their open clusters in gold and binaries retained in their open clusters in purple. We find that most Gaia BHs are retained for at least 100 Myr in the open clusters in which they formed. However, since many systems will likely still escape at later times, this does not necessarily indicate that (if BH-LC systems preferentially form via the dynamical channel) we should expect to discover many Gaia BHs in open clusters in future \textit{Gaia} data releases. Moreover, even if Gaia BHs are retained in open clusters, they could be difficult to detect due to the effects of crowding.

In all panels, there is a steep drop off in the period distribution at $\sim10^{3.5}$ days, which reflects the fact that the input period distribution from \citet{sana_binary_2012} only extends to $10^{3.5}$ days. However, Figure~\ref{fig:dyn_prop_plot} shows that most of the binaries are exchanged, not primordial. This may appear counter-intuitive, since one may expect exchanges to substantially alter the input period distribution. However, the models from \citet{di_carlo_young_2024} predict that the post-exchange period is usually similar to or shorter than the pre-exchange period \citep[see also][]{Sigurdsson1993}, so the pre-exchange period distribution, and thus the assumed binary evolution, is actually quite important. 

With this in mind, Figure~\ref{fig:compare_pop_plot} compares the period-eccentricity relations for the populations formed via IBE in the two models. The truncated initial period distribution and larger common envelope $\alpha$ adopted by \citet{di_carlo_young_2024} respectively cause the model to predict fewer long-period ($P_{\rm orb}>10^4$ d)  binaries and wider post-interaction binaries than the model of \citet{chawla_gaia_2022}, including a significant number with $P_{\rm orb} = 10^2$--$10^4$ d.

Since \textit{Gaia} astrometry is most sensitive to binaries with orbital periods of $10^2$--$10^4$ d, and \citet{di_carlo_young_2024} use the same binary evolution model for both isolated and dynamical formation channels, Figure~\ref{fig:compare_pop_plot} may partially explain why Gaia BHs are easier to discover in the dynamically assembled BH-LC population of \citet{di_carlo_young_2024}: the model's binary evolution parameters result in more BH-LC binaries with intermediate periods, and some memory of these periods is retained after exchange interactions. 


\section{Conclusion} \label{sec:conclusion}

We have applied the generative model of \citet{2024OJAp....7E.100E} to model populations of black hole (BH)-luminous companion (LC) systems formed via isolated binary evolution (IBE) \citep{chawla_gaia_2022} or dynamical interactions \citep{di_carlo_young_2024} to predict the number of Gaia BHs detectable in \textit{Gaia} DR3, DR4, and DR5. Our modeling improves on previous work by simulating astrometric measurements at the epoch level (Figure~\ref{fig:ex_astro_fits}) and replicating the \textit{Gaia} DR3 astrometric cascade. We summarize our results below.

\begin{itemize}
    \item The period distribution of the IBE population is bimodal, with a gap at $\sim10^2$--$10^4$ days, where \textit{Gaia} is most sensitive. In contrast, the period distribution of the dynamical population (which includes many binaries in this period range) is roughly log-uniform up to periods of $\sim10^{3.5}$ days (Figures~\ref{fig:intrinsic_props} and \ref{fig:both_sma}). The IBE population features LCs that are brighter and more massive relative to the dynamical population (Figure~\ref{fig:intrinsic_props}). As a result, the spatial distribution of the IBE population is predicted to be clumpier than the dynamical population, leading to more scatter in the number of detectable Gaia BHs across Milky Way (MW) realizations (Figure~\ref{fig:mw_dist_plot}). 
    \item The IBE formation channel predicts that no Gaia BHs are detectable in DR3 in the vast majority of MW realizations, with the few detectable LCs all being significantly more luminous than the Sun. On the other hand, the median number (across 100 MW realizations) of dynamically formed Gaia BHs detected in DR3 is $17^{+6}_{-5}$. Based on a comparison to the number of Gaia BHs already discovered in DR3, we conclude that the IBE formation channel underestimates the number of potentially detectable BH-LC systems, while the dynamical formation channel overestimates that number by a factor of $\sim 8$ (Figure~\ref{fig:dr3_bar_plot}).
    \item Longer observing baselines in DR4 and DR5 will allow for the detection of more Gaia BHs at larger distances (Figures~\ref{fig:isolated_scatter} and \ref{fig:dynamic_scatter}). This is because longer baselines provide sensitivity to binaries with longer orbital periods, which have larger physical orbits and can consequently be detected to larger distances with \textit{Gaia} astrometry. Thus, the search volume increases rapidly as the time baseline increases. 
    
    Adopting simple detectability cuts of $\varpi / \sigma_{\varpi} > 5$ and $a_0 / \sigma_{a_0} > 5$ and calibrating to DR3, we predict that $30^{+2}_{-3}$ Gaia BHs will be detectable in DR4, with that number increasing to $45^{+4}_{-5}$ in DR5 (Figure~\ref{fig:bar_plot}). Several of our assumptions are conservative, so these are probably lower limits. Adopting the stricter detectability cuts used in DR3 decreases the amount of BH-LC systems detectable in future data releases by a factor of $\sim 2$ (Figure~\ref{fig:bar_plot}). Based on our results, we estimate that 1 in $10^7$ stars in the MW should orbit a BH in an au-scale orbit, and that about 1 in 1000 of these systems will receive an astrometric orbital solution in \textit{Gaia} DR4.
    \item Detected Gaia BHs tend to have orbital periods of $10^2$--$10^4$ days. They fall in the ``period gap'' in-between the wide, non-interacting binaries and close, post-common envelope binaries formed via IBE (Figure~\ref{fig:isolated_scatter}), but can be formed in larger numbers via dynamical assembly (Figure~\ref{fig:dynamic_scatter}). The detected BH-LC systems, on average, are brighter and contain more more massive luminous stars than the underlying population (Figures~\ref{fig:isolated_corner} and \ref{fig:dynamic_corner}). BH-LC systems formed via IBE tend to have circular orbits, while those formed via dynamical interactions have support at all eccentricities (Figures~\ref{fig:isolated_corner} and \ref{fig:dynamic_corner}).
    \item We fit 7-parameter and 9-parameter acceleration solutions to the dynamically formed BH-LC systems with RUWE $> 1.4$ in DR4. Applying the cuts imposed upon acceleration solutions in DR3, we find that no 7-parameter solutions and only one 9-parameter solution find their way into the final catalog, suggesting that acceleration solutions do not represent a promising method of searching for dormant BHs (Figure \ref{fig:acc_plot}). Computing RUWE for all dynamically formed BH-LC systems, we find that systems that eventually received an astrometric binary solution have large RUWE values and lie at close distances relative to the underlying population (Figure \ref{fig:ruwe_plot}). There are many false negatives at RUWE values of $2$--$5$ that were rejected because they did not pass DR3 cuts designed to remove spurious astrometric solutions, implying that RUWE-based searches can uncover Gaia BHs that would otherwise remain undetected. 
    \item To understand why the two formation models predict such disparate results, we compare the physical properties of the intrinsic populations (Figures~ \ref{fig:dyn_prop_plot} and \ref{fig:compare_pop_plot}). We find the difference in the assumed initial period distribution and binary evolution modeling is likely an important factor, since the dynamically formed BH binary population still bears imprints of the primordial binary population. We conclude that dynamical simulations exploring a wider variety of primordial binary populations and binary evolution physics are needed.
\end{itemize}

Discoveries of Gaia BHs in future data releases will validate the  predictions of this work.

\begin{acknowledgments}

We thank the referee for useful comments. We thank Richard O'Shaughnessy and members of the Caltech--Carnegie dormant black hole task force for useful discussions. This research was supported by NSF grants AST-2307232 and AST-2310362. CR acknowledges support from the Alfred P. Sloan Foundation and the David and Lucile Packard Foundation. This work has made use of data from the European Space Agency (ESA) mission
{\it Gaia} (\url{https://www.cosmos.esa.int/gaia}), processed by the {\it Gaia}
Data Processing and Analysis Consortium (DPAC,
\url{https://www.cosmos.esa.int/web/gaia/dpac/consortium}). Funding for the DPAC
has been provided by national institutions, in particular the institutions
participating in the {\it Gaia} Multilateral Agreement.

\end{acknowledgments}

%



\software{astropy \citep{2013A&A...558A..33A,2018AJ....156..123A}}





\bibliography{bibliography}{}

\begin{thebibliography}{}
\expandafter\ifx\csname natexlab\endcsname\relax\def\natexlab#1{#1}\fi
\providecommand{\url}[1]{\href{#1}{#1}}
\providecommand{\dodoi}[1]{doi:~\href{http://doi.org/#1}{\nolinkurl{#1}}}
\providecommand{\doeprint}[1]{\href{http://ascl.net/#1}{\nolinkurl{http://ascl.net/#1}}}
\providecommand{\doarXiv}[1]{\href{https://arxiv.org/abs/#1}{\nolinkurl{https://arxiv.org/abs/#1}}}

\bibitem[{Andrews {et~al.}(2019)Andrews, Breivik, \& Chatterjee}]{andrews_weighing_2019}
Andrews, J.~J., Breivik, K., \& Chatterjee, S. 2019, The Astrophysical Journal, 886, 68, \dodoi{10.3847/1538-4357/ab441f}

\bibitem[{{Astropy Collaboration} {et~al.}(2013){Astropy Collaboration}, {Robitaille}, {Tollerud}, {Greenfield}, {Droettboom}, {Bray}, {Aldcroft}, {Davis}, {Ginsburg}, {Price-Whelan}, {Kerzendorf}, {Conley}, {Crighton}, {Barbary}, {Muna}, {Ferguson}, {Grollier}, {Parikh}, {Nair}, {Unther}, {Deil}, {Woillez}, {Conseil}, {Kramer}, {Turner}, {Singer}, {Fox}, {Weaver}, {Zabalza}, {Edwards}, {Azalee Bostroem}, {Burke}, {Casey}, {Crawford}, {Dencheva}, {Ely}, {Jenness}, {Labrie}, {Lim}, {Pierfederici}, {Pontzen}, {Ptak}, {Refsdal}, {Servillat}, \& {Streicher}}]{2013A&A...558A..33A}
{Astropy Collaboration}, {Robitaille}, T.~P., {Tollerud}, E.~J., {et~al.} 2013, \aap, 558, A33, \dodoi{10.1051/0004-6361/201322068}

\bibitem[{{Astropy Collaboration} {et~al.}(2018){Astropy Collaboration}, {Price-Whelan}, {Sip{\H{o}}cz}, {G{\"u}nther}, {Lim}, {Crawford}, {Conseil}, {Shupe}, {Craig}, {Dencheva}, {Ginsburg}, {VanderPlas}, {Bradley}, {P{\'e}rez-Su{\'a}rez}, {de Val-Borro}, {Aldcroft}, {Cruz}, {Robitaille}, {Tollerud}, {Ardelean}, {Babej}, {Bach}, {Bachetti}, {Bakanov}, {Bamford}, {Barentsen}, {Barmby}, {Baumbach}, {Berry}, {Biscani}, {Boquien}, {Bostroem}, {Bouma}, {Brammer}, {Bray}, {Breytenbach}, {Buddelmeijer}, {Burke}, {Calderone}, {Cano Rodr{\'\i}guez}, {Cara}, {Cardoso}, {Cheedella}, {Copin}, {Corrales}, {Crichton}, {D'Avella}, {Deil}, {Depagne}, {Dietrich}, {Donath}, {Droettboom}, {Earl}, {Erben}, {Fabbro}, {Ferreira}, {Finethy}, {Fox}, {Garrison}, {Gibbons}, {Goldstein}, {Gommers}, {Greco}, {Greenfield}, {Groener}, {Grollier}, {Hagen}, {Hirst}, {Homeier}, {Horton}, {Hosseinzadeh}, {Hu}, {Hunkeler}, {Ivezi{\'c}}, {Jain}, {Jenness}, {Kanarek}, {Kendrew}, {Kern}, {Kerzendorf}, {Khvalko}, {King}, {Kirkby}, {Kulkarni},
  {Kumar}, {Lee}, {Lenz}, {Littlefair}, {Ma}, {Macleod}, {Mastropietro}, {McCully}, {Montagnac}, {Morris}, {Mueller}, {Mumford}, {Muna}, {Murphy}, {Nelson}, {Nguyen}, {Ninan}, {N{\"o}the}, {Ogaz}, {Oh}, {Parejko}, {Parley}, {Pascual}, {Patil}, {Patil}, {Plunkett}, {Prochaska}, {Rastogi}, {Reddy Janga}, {Sabater}, {Sakurikar}, {Seifert}, {Sherbert}, {Sherwood-Taylor}, {Shih}, {Sick}, {Silbiger}, {Singanamalla}, {Singer}, {Sladen}, {Sooley}, {Sornarajah}, {Streicher}, {Teuben}, {Thomas}, {Tremblay}, {Turner}, {Terr{\'o}n}, {van Kerkwijk}, {de la Vega}, {Watkins}, {Weaver}, {Whitmore}, {Woillez}, {Zabalza}, \& {Astropy Contributors}}]{2018AJ....156..123A}
{Astropy Collaboration}, {Price-Whelan}, A.~M., {Sip{\H{o}}cz}, B.~M., {et~al.} 2018, \aj, 156, 123, \dodoi{10.3847/1538-3881/aabc4f}

\bibitem[{{Bovy} {et~al.}(2016){Bovy}, {Rix}, {Green}, {Schlafly}, \& {Finkbeiner}}]{bovy_2016}
{Bovy}, J., {Rix}, H.-W., {Green}, G.~M., {Schlafly}, E.~F., \& {Finkbeiner}, D.~P. 2016, \apj, 818, 130, \dodoi{10.3847/0004-637X/818/2/130}

\bibitem[{Breivik {et~al.}(2017)Breivik, Chatterjee, \& Larson}]{breivik_revealing_2017}
Breivik, K., Chatterjee, S., \& Larson, S.~L. 2017, The Astrophysical Journal, 850, L13, \dodoi{10.3847/2041-8213/aa97d5}

\bibitem[{{Breivik} {et~al.}(2020){Breivik}, {Coughlin}, {Zevin}, {Rodriguez}, {Kremer}, {Ye}, {Andrews}, {Kurkowski}, {Digman}, {Larson}, \& {Rasio}}]{breivik_cosmic_2020}
{Breivik}, K., {Coughlin}, S., {Zevin}, M., {et~al.} 2020, \apj, 898, 71, \dodoi{10.3847/1538-4357/ab9d85}

\bibitem[{Chakrabarti {et~al.}(2023)Chakrabarti, Simon, Craig, Reggiani, Brandt, Guhathakurta, Dalba, Kirby, Chang, Hey, Savino, Geha, \& Thompson}]{chakrabarti_noninteracting_2023}
Chakrabarti, S., Simon, J.~D., Craig, P.~A., {et~al.} 2023, The Astronomical Journal, 166, 6, \dodoi{10.3847/1538-3881/accf21}

\bibitem[{Chawla {et~al.}(2022)Chawla, Chatterjee, Breivik, Moorthy, Andrews, \& Sanderson}]{chawla_gaia_2022}
Chawla, C., Chatterjee, S., Breivik, K., {et~al.} 2022, The Astrophysical Journal, 931, 107, \dodoi{10.3847/1538-4357/ac60a5}

\bibitem[{{Claeys} {et~al.}(2014){Claeys}, {Pols}, {Izzard}, {Vink}, \& {Verbunt}}]{claeys_2014}
{Claeys}, J.~S.~W., {Pols}, O.~R., {Izzard}, R.~G., {Vink}, J., \& {Verbunt}, F.~W.~M. 2014, \aap, 563, A83, \dodoi{10.1051/0004-6361/201322714}

\bibitem[{Di~Carlo {et~al.}(2024)Di~Carlo, Agrawal, Rodriguez, \& Breivik}]{di_carlo_young_2024}
Di~Carlo, U.~N., Agrawal, P., Rodriguez, C.~L., \& Breivik, K. 2024, The Astrophysical Journal, 965, 22, \dodoi{10.3847/1538-4357/ad2f2c}

\bibitem[{{Di Carlo} {et~al.}(2020){Di Carlo}, {Mapelli}, {Giacobbo}, {Spera}, {Bouffanais}, {Rastello}, {Santoliquido}, {Pasquato}, {Ballone}, {Trani}, {Torniamenti}, \& {Haardt}}]{di_carlo_2020}
{Di Carlo}, U.~N., {Mapelli}, M., {Giacobbo}, N., {et~al.} 2020, \mnras, 498, 495, \dodoi{10.1093/mnras/staa2286}

\bibitem[{{Drimmel} {et~al.}(2003){Drimmel}, {Cabrera-Lavers}, \& {L{\'o}pez-Corredoira}}]{drimmel_2003}
{Drimmel}, R., {Cabrera-Lavers}, A., \& {L{\'o}pez-Corredoira}, M. 2003, \aap, 409, 205, \dodoi{10.1051/0004-6361:20031070}

\bibitem[{El-Badry(2024)}]{el-badry_gaias_2024}
El-Badry, K. 2024, New Astronomy Reviews, 98, 101694, \dodoi{10.1016/j.newar.2024.101694}

\bibitem[{{El-Badry} {et~al.}(2024){El-Badry}, {Lam}, {Holl}, {Halbwachs}, {Rix}, {Mazeh}, \& {Shahaf}}]{2024OJAp....7E.100E}
{El-Badry}, K., {Lam}, C., {Holl}, B., {et~al.} 2024, The Open Journal of Astrophysics, 7, 100, \dodoi{10.33232/001c.125461}

\bibitem[{El-Badry {et~al.}(2023{\natexlab{a}})El-Badry, Rix, Quataert, Howard, Isaacson, Fuller, Hawkins, Breivik, Wong, Rodriguez, Conroy, Shahaf, Mazeh, Arenou, Burdge, Bashi, Faigler, Weisz, Seeburger, Almada~Monter, \& Wojno}]{el-badry_sun-like_2023}
El-Badry, K., Rix, H.-W., Quataert, E., {et~al.} 2023{\natexlab{a}}, Monthly Notices of the Royal Astronomical Society, 518, 1057, \dodoi{10.1093/mnras/stac3140}

\bibitem[{El-Badry {et~al.}(2023{\natexlab{b}})El-Badry, Rix, Cendes, Rodriguez, Conroy, Quataert, Hawkins, Zari, Hobson, Breivik, Rau, Berger, Shahaf, Seeburger, Burdge, Latham, Buchhave, Bieryla, Bashi, Mazeh, \& Faigler}]{el-badry_red_2023}
El-Badry, K., Rix, H.-W., Cendes, Y., {et~al.} 2023{\natexlab{b}}, Monthly Notices of the Royal Astronomical Society, 521, 4323, \dodoi{10.1093/mnras/stad799}

\bibitem[{El-Badry {et~al.}(2024)El-Badry, Rix, Latham, Shahaf, Mazeh, Bieryla, Buchhave, Andrae, Yamaguchi, Isaacson, Howard, Savino, \& Ilyin}]{el-badry_population_2024}
El-Badry, K., Rix, H.-W., Latham, D.~W., {et~al.} 2024, The Open Journal of Astrophysics, 7, 58, \dodoi{10.33232/001c.121261}

\bibitem[{{Fantoccoli} {et~al.}(2025){Fantoccoli}, {Barber}, {Dosopoulou}, {Chattopadhyay}, \& {Antonini}}]{fantoccoli_2024}
{Fantoccoli}, F., {Barber}, J., {Dosopoulou}, F., {Chattopadhyay}, D., \& {Antonini}, F. 2025, \mnras, \dodoi{10.1093/mnras/staf303}

\bibitem[{{Fryer} {et~al.}(2012){Fryer}, {Belczynski}, {Wiktorowicz}, {Dominik}, {Kalogera}, \& {Holz}}]{fryer_2012}
{Fryer}, C.~L., {Belczynski}, K., {Wiktorowicz}, G., {et~al.} 2012, \apj, 749, 91, \dodoi{10.1088/0004-637X/749/1/91}

\bibitem[{{Gaia Collaboration} {et~al.}(2016){Gaia Collaboration}, Prusti, de~Bruijne, Brown, Vallenari, Babusiaux, Bailer-Jones, Bastian, Biermann, Evans, Eyer, Jansen, Jordi, Klioner, Lammers, Lindegren, Luri, Mignard, Milligan, Panem, Poinsignon, Pourbaix, Randich, Sarri, Sartoretti, Siddiqui, Soubiran, Valette, van Leeuwen, Walton, Aerts, Arenou, Cropper, Drimmel, Høg, Katz, Lattanzi, O'Mullane, Grebel, Holland, Huc, Passot, Bramante, Cacciari, Castañeda, Chaoul, Cheek, De~Angeli, Fabricius, Guerra, Hernández, Jean-Antoine-Piccolo, Masana, Messineo, Mowlavi, Nienartowicz, Ordóñez-Blanco, Panuzzo, Portell, Richards, Riello, Seabroke, Tanga, Thévenin, Torra, Els, Gracia-Abril, Comoretto, Garcia-Reinaldos, Lock, Mercier, Altmann, Andrae, Astraatmadja, Bellas-Velidis, Benson, Berthier, Blomme, Busso, Carry, Cellino, Clementini, Cowell, Creevey, Cuypers, Davidson, De~Ridder, de~Torres, Delchambre, Dell'Oro, Ducourant, Frémat, García-Torres, Gosset, Halbwachs, Hambly, Harrison, Hauser, Hestroffer,
  Hodgkin, Huckle, Hutton, Jasniewicz, Jordan, Kontizas, Korn, Lanzafame, Manteiga, Moitinho, Muinonen, Osinde, Pancino, Pauwels, Petit, Recio-Blanco, Robin, Sarro, Siopis, Smith, Smith, Sozzetti, Thuillot, van Reeven, Viala, Abbas, Abreu~Aramburu, Accart, Aguado, Allan, Allasia, Altavilla, Álvarez, Alves, Anderson, Andrei, Anglada~Varela, Antiche, Antoja, Antón, Arcay, Atzei, Ayache, Bach, Baker, Balaguer-Núñez, Barache, Barata, Barbier, Barblan, Baroni, Barrado~y Navascués, Barros, Barstow, Becciani, Bellazzini, Bellei, Bello~García, Belokurov, Bendjoya, Berihuete, Bianchi, Bienaymé, Billebaud, Blagorodnova, Blanco-Cuaresma, Boch, Bombrun, Borrachero, Bouquillon, Bourda, Bouy, Bragaglia, Breddels, Brouillet, Brüsemeister, Bucciarelli, Budnik, Burgess, Burgon, Burlacu, Busonero, Buzzi, Caffau, Cambras, Campbell, Cancelliere, Cantat-Gaudin, Carlucci, Carrasco, Castellani, Charlot, Charnas, Charvet, Chassat, Chiavassa, Clotet, Cocozza, Collins, Collins, Costigan, Crifo, Cross, Crosta, Crowley, Dafonte,
  Damerdji, Dapergolas, David, David, De~Cat, de~Felice, de~Laverny, De~Luise, De~March, de~Martino, de~Souza, Debosscher, del Pozo, Delbo, Delgado, Delgado, di~Marco, Di~Matteo, Diakite, Distefano, Dolding, Dos~Anjos, Drazinos, Durán, Dzigan, Ecale, Edvardsson, Enke, Erdmann, Escolar, Espina, Evans, Eynard~Bontemps, Fabre, Fabrizio, Faigler, Falcão, Farràs~Casas, Faye, Federici, Fedorets, Fernández-Hernández, Fernique, Fienga, Figueras, Filippi, Findeisen, Fonti, Fouesneau, Fraile, Fraser, Fuchs, Furnell, Gai, Galleti, Galluccio, Garabato, García-Sedano, Garé, Garofalo, Garralda, Gavras, Gerssen, Geyer, Gilmore, Girona, Giuffrida, Gomes, González-Marcos, González-Núñez, González-Vidal, Granvik, Guerrier, Guillout, Guiraud, Gúrpide, Gutiérrez-Sánchez, Guy, Haigron, Hatzidimitriou, Haywood, Heiter, Helmi, Hobbs, Hofmann, Holl, Holland, Hunt, Hypki, Icardi, Irwin, Jevardat~de Fombelle, Jofré, Jonker, Jorissen, Julbe, Karampelas, Kochoska, Kohley, Kolenberg, Kontizas, Koposov, Kordopatis,
  Koubsky, Kowalczyk, Krone-Martins, Kudryashova, Kull, Bachchan, Lacoste-Seris, Lanza, Lavigne, Le~Poncin-Lafitte, Lebreton, Lebzelter, Leccia, Leclerc, Lecoeur-Taibi, Lemaitre, Lenhardt, Leroux, Liao, Licata, Lindstrøm, Lister, Livanou, Lobel, Löffler, López, Lopez-Lozano, Lorenz, Loureiro, MacDonald, Magalhães~Fernandes, Managau, Mann, Mantelet, Marchal, Marchant, Marconi, Marie, Marinoni, Marrese, Marschalkó, Marshall, Martín-Fleitas, Martino, Mary, Matijevič, Mazeh, McMillan, Messina, Mestre, Michalik, Millar, Miranda, Molina, Molinaro, Molinaro, Molnár, Moniez, Montegriffo, Monteiro, Mor, Mora, Morbidelli, Morel, Morgenthaler, Morley, Morris, Mulone, Muraveva, Musella, Narbonne, Nelemans, Nicastro, Noval, Ordénovic, Ordieres-Meré, Osborne, Pagani, Pagano, Pailler, Palacin, Palaversa, Parsons, Paulsen, Pecoraro, Pedrosa, Pentikäinen, Pereira, Pichon, Piersimoni, Pineau, Plachy, Plum, Poujoulet, Prša, Pulone, Ragaini, Rago, Rambaux, Ramos-Lerate, Ranalli, Rauw, Read, Regibo, Renk, Reylé,
  Ribeiro, Rimoldini, Ripepi, Riva, Rixon, Roelens, Romero-Gómez, Rowell, Royer, Rudolph, Ruiz-Dern, Sadowski, Sagristà~Sellés, Sahlmann, Salgado, Salguero, Sarasso, Savietto, Schnorhk, Schultheis, Sciacca, Segol, Segovia, Segransan, Serpell, Shih, Smareglia, Smart, Smith, Solano, Solitro, Sordo, Soria~Nieto, Souchay, Spagna, Spoto, Stampa, Steele, Steidelmüller, Stephenson, Stoev, Suess, Süveges, Surdej, Szabados, Szegedi-Elek, Tapiador, Taris, Tauran, Taylor, Teixeira, Terrett, Tingley, Trager, Turon, Ulla, Utrilla, Valentini, van Elteren, Van~Hemelryck, van Leeuwen, Varadi, Vecchiato, Veljanoski, Via, Vicente, Vogt, Voss, Votruba, Voutsinas, Walmsley, Weiler, Weingrill, Werner, Wevers, Whitehead, Wyrzykowski, Yoldas, Žerjal, Zucker, Zurbach, Zwitter, Alecu, Allen, Allende~Prieto, Amorim, Anglada-Escudé, Arsenijevic, Azaz, Balm, Beck, Bernstein, Bigot, Bijaoui, Blasco, Bonfigli, Bono, Boudreault, Bressan, Brown, Brunet, Bunclark, Buonanno, Butkevich, Carret, Carrion, Chemin, Chéreau, Corcione,
  Darmigny, de~Boer, de~Teodoro, de~Zeeuw, Delle~Luche, Domingues, Dubath, Fodor, Frézouls, Fries, Fustes, Fyfe, Gallardo, Gallegos, Gardiol, Gebran, Gomboc, Gómez, Grux, Gueguen, Heyrovsky, Hoar, Iannicola, Isasi~Parache, Janotto, Joliet, Jonckheere, Keil, Kim, Klagyivik, Klar, Knude, Kochukhov, Kolka, Kos, Kutka, Lainey, LeBouquin, Liu, Loreggia, Makarov, Marseille, Martayan, Martinez-Rubi, Massart, Meynadier, Mignot, Munari, Nguyen, Nordlander, Ocvirk, O'Flaherty, Olias~Sanz, Ortiz, Osorio, Oszkiewicz, Ouzounis, Palmer, Park, Pasquato, Peltzer, Peralta, Péturaud, Pieniluoma, Pigozzi, Poels, Prat, Prod'homme, Raison, Rebordao, Risquez, Rocca-Volmerange, Rosen, Ruiz-Fuertes, Russo, Sembay, Serraller~Vizcaino, Short, Siebert, Silva, Sinachopoulos, Slezak, Soffel, Sosnowska, Straižys, ter Linden, Terrell, Theil, Tiede, Troisi, Tsalmantza, Tur, Vaccari, Vachier, Valles, Van~Hamme, Veltz, Virtanen, Wallut, Wichmann, Wilkinson, Ziaeepour, \& Zschocke}]{gaia_collaboration_gaia_2016}
{Gaia Collaboration}, Prusti, T., de~Bruijne, J. H.~J., {et~al.} 2016, Astronomy and Astrophysics, 595, A1, \dodoi{10.1051/0004-6361/201629272}

\bibitem[{{Gaia Collaboration} {et~al.}(2023{\natexlab{a}}){Gaia Collaboration}, Vallenari, Brown, Prusti, de~Bruijne, Arenou, Babusiaux, Biermann, Creevey, Ducourant, Evans, Eyer, Guerra, Hutton, Jordi, Klioner, Lammers, Lindegren, Luri, Mignard, Panem, Pourbaix, Randich, Sartoretti, Soubiran, Tanga, Walton, Bailer-Jones, Bastian, Drimmel, Jansen, Katz, Lattanzi, van Leeuwen, Bakker, Cacciari, Castañeda, De~Angeli, Fabricius, Fouesneau, Frémat, Galluccio, Guerrier, Heiter, Masana, Messineo, Mowlavi, Nicolas, Nienartowicz, Pailler, Panuzzo, Riclet, Roux, Seabroke, Sordo, Thévenin, Gracia-Abril, Portell, Teyssier, Altmann, Andrae, Audard, Bellas-Velidis, Benson, Berthier, Blomme, Burgess, Busonero, Busso, Cánovas, Carry, Cellino, Cheek, Clementini, Damerdji, Davidson, de~Teodoro, Nuñez~Campos, Delchambre, Dell'Oro, Esquej, Fernández-Hernández, Fraile, Garabato, García-Lario, Gosset, Haigron, Halbwachs, Hambly, Harrison, Hernández, Hestroffer, Hodgkin, Holl, Janßen, Jevardat~de Fombelle, Jordan,
  Krone-Martins, Lanzafame, Löffler, Marchal, Marrese, Moitinho, Muinonen, Osborne, Pancino, Pauwels, Recio-Blanco, Reylé, Riello, Rimoldini, Roegiers, Rybizki, Sarro, Siopis, Smith, Sozzetti, Utrilla, van Leeuwen, Abbas, Ábrahám, Abreu~Aramburu, Aerts, Aguado, Ajaj, Aldea-Montero, Altavilla, Álvarez, Alves, Anders, Anderson, Anglada~Varela, Antoja, Baines, Baker, Balaguer-Núñez, Balbinot, Balog, Barache, Barbato, Barros, Barstow, Bartolomé, Bassilana, Bauchet, Becciani, Bellazzini, Berihuete, Bernet, Bertone, Bianchi, Binnenfeld, Blanco-Cuaresma, Blazere, Boch, Bombrun, Bossini, Bouquillon, Bragaglia, Bramante, Breedt, Bressan, Brouillet, Brugaletta, Bucciarelli, Burlacu, Butkevich, Buzzi, Caffau, Cancelliere, Cantat-Gaudin, Carballo, Carlucci, Carnerero, Carrasco, Casamiquela, Castellani, Castro-Ginard, Chaoul, Charlot, Chemin, Chiaramida, Chiavassa, Chornay, Comoretto, Contursi, Cooper, Cornez, Cowell, Crifo, Cropper, Crosta, Crowley, Dafonte, Dapergolas, David, David, de~Laverny, De~Luise,
  De~March, De~Ridder, de~Souza, de~Torres, del Peloso, del Pozo, Delbo, Delgado, Delisle, Demouchy, Dharmawardena, Di~Matteo, Diakite, Diener, Distefano, Dolding, Edvardsson, Enke, Fabre, Fabrizio, Faigler, Fedorets, Fernique, Fienga, Figueras, Fournier, Fouron, Fragkoudi, Gai, Garcia-Gutierrez, Garcia-Reinaldos, García-Torres, Garofalo, Gavel, Gavras, Gerlach, Geyer, Giacobbe, Gilmore, Girona, Giuffrida, Gomel, Gomez, González-Núñez, González-Santamaría, González-Vidal, Granvik, Guillout, Guiraud, Gutiérrez-Sánchez, Guy, Hatzidimitriou, Hauser, Haywood, Helmer, Helmi, Sarmiento, Hidalgo, Hilger, Hładczuk, Hobbs, Holland, Huckle, Jardine, Jasniewicz, Jean-Antoine~Piccolo, Jiménez-Arranz, Jorissen, Juaristi~Campillo, Julbe, Karbevska, Kervella, Khanna, Kontizas, Kordopatis, Korn, Kóspál, Kostrzewa-Rutkowska, Kruszyńska, Kun, Laizeau, Lambert, Lanza, Lasne, Le~Campion, Lebreton, Lebzelter, Leccia, Leclerc, Lecoeur-Taibi, Liao, Licata, Lindstrøm, Lister, Livanou, Lobel, Lorca, Loup,
  Madrero~Pardo, Magdaleno~Romeo, Managau, Mann, Manteiga, Marchant, Marconi, Marcos, Marcos~Santos, Marín~Pina, Marinoni, Marocco, Marshall, Martin~Polo, Martín-Fleitas, Marton, Mary, Masip, Massari, Mastrobuono-Battisti, Mazeh, McMillan, Messina, Michalik, Millar, Mints, Molina, Molinaro, Molnár, Monari, Monguió, Montegriffo, Montero, Mor, Mora, Morbidelli, Morel, Morris, Muraveva, Murphy, Musella, Nagy, Noval, Ocaña, Ogden, Ordenovic, Osinde, Pagani, Pagano, Palaversa, Palicio, Pallas-Quintela, Panahi, Payne-Wardenaar, Peñalosa~Esteller, Penttilä, Pichon, Piersimoni, Pineau, Plachy, Plum, Poggio, Prša, Pulone, Racero, Ragaini, Rainer, Raiteri, Rambaux, Ramos, Ramos-Lerate, Re~Fiorentin, Regibo, Richards, Rios~Diaz, Ripepi, Riva, Rix, Rixon, Robichon, Robin, Robin, Roelens, Rogues, Rohrbasser, Romero-Gómez, Rowell, Royer, Ruz~Mieres, Rybicki, Sadowski, Sáez~Núñez, Sagristà~Sellés, Sahlmann, Salguero, Samaras, Sanchez~Gimenez, Sanna, Santoveña, Sarasso, Schultheis, Sciacca, Segol, Segovia,
  Ségransan, Semeux, Shahaf, Siddiqui, Siebert, Siltala, Silvelo, Slezak, Slezak, Smart, Snaith, Solano, Solitro, Souami, Souchay, Spagna, Spina, Spoto, Steele, Steidelmüller, Stephenson, Süveges, Surdej, Szabados, Szegedi-Elek, Taris, Taylor, Teixeira, Tolomei, Tonello, Torra, Torra, Torralba~Elipe, Trabucchi, Tsounis, Turon, Ulla, Unger, Vaillant, van Dillen, van Reeven, Vanel, Vecchiato, Viala, Vicente, Voutsinas, Weiler, Wevers, Wyrzykowski, Yoldas, Yvard, Zhao, Zorec, Zucker, \& Zwitter}]{gaia_collaboration_gaia_2023}
{Gaia Collaboration}, Vallenari, A., Brown, A. G.~A., {et~al.} 2023{\natexlab{a}}, Astronomy and Astrophysics, 674, A1, \dodoi{10.1051/0004-6361/202243940}

\bibitem[{{Gaia Collaboration} {et~al.}(2023{\natexlab{b}}){Gaia Collaboration}, Arenou, Babusiaux, Barstow, Faigler, Jorissen, Kervella, Mazeh, Mowlavi, Panuzzo, Sahlmann, Shahaf, Sozzetti, Bauchet, Damerdji, Gavras, Giacobbe, Gosset, Halbwachs, Holl, Lattanzi, Leclerc, Morel, Pourbaix, Re~Fiorentin, Sadowski, Ségransan, Siopis, Teyssier, Zwitter, Planquart, Brown, Vallenari, Prusti, de~Bruijne, Biermann, Creevey, Ducourant, Evans, Eyer, Guerra, Hutton, Jordi, Klioner, Lammers, Lindegren, Luri, Mignard, Panem, Randich, Sartoretti, Soubiran, Tanga, Walton, Bailer-Jones, Bastian, Drimmel, Jansen, Katz, van Leeuwen, Bakker, Cacciari, Castañeda, De~Angeli, Fabricius, Fouesneau, Frémat, Galluccio, Guerrier, Heiter, Masana, Messineo, Nicolas, Nienartowicz, Pailler, Riclet, Roux, Seabroke, Sordo, Thévenin, Gracia-Abril, Portell, Altmann, Andrae, Audard, Bellas-Velidis, Benson, Berthier, Blomme, Burgess, Busonero, Busso, Cánovas, Carry, Cellino, Cheek, Clementini, Davidson, de~Teodoro, Nuñez~Campos,
  Delchambre, Dell'Oro, Esquej, Fernández-Hernández, Fraile, Garabato, García-Lario, Haigron, Hambly, Harrison, Hernández, Hestroffer, Hodgkin, Janßen, Jevardat~de Fombelle, Jordan, Krone-Martins, Lanzafame, Löffler, Marchal, Marrese, Moitinho, Muinonen, Osborne, Pancino, Pauwels, Recio-Blanco, Reylé, Riello, Rimoldini, Roegiers, Rybizki, Sarro, Smith, Utrilla, van Leeuwen, Abbas, Ábrahám, Abreu~Aramburu, Aerts, Aguado, Ajaj, Aldea-Montero, Altavilla, Álvarez, Alves, Anders, Anderson, Anglada~Varela, Antoja, Baines, Baker, Balaguer-Núñez, Balbinot, Balog, Barache, Barbato, Barros, Bartolomé, Bassilana, Becciani, Bellazzini, Berihuete, Bernet, Bertone, Bianchi, Binnenfeld, Blanco-Cuaresma, Blazere, Boch, Bombrun, Bossini, Bouquillon, Bragaglia, Bramante, Breedt, Bressan, Brouillet, Brugaletta, Bucciarelli, Burlacu, Butkevich, Buzzi, Caffau, Cancelliere, Cantat-Gaudin, Carballo, Carlucci, Carnerero, Carrasco, Casamiquela, Castellani, Castro-Ginard, Chaoul, Charlot, Chemin, Chiaramida, Chiavassa,
  Chornay, Comoretto, Contursi, Cooper, Cornez, Cowell, Crifo, Cropper, Crosta, Crowley, Dafonte, Dapergolas, David, de~Laverny, De~Luise, De~March, De~Ridder, de~Souza, de~Torres, del Peloso, del Pozo, Delbo, Delgado, Delisle, Demouchy, Dharmawardena, Diakite, Diener, Distefano, Dolding, Enke, Fabre, Fabrizio, Fedorets, Fernique, Figueras, Fournier, Fouron, Fragkoudi, Gai, Garcia-Gutierrez, Garcia-Reinaldos, García-Torres, Garofalo, Gavel, Gerlach, Geyer, Gilmore, Girona, Giuffrida, Gomel, Gomez, González-Núñez, González-Santamaría, González-Vidal, Granvik, Guillout, Guiraud, Gutiérrez-Sánchez, Guy, Hatzidimitriou, Hauser, Haywood, Helmer, Helmi, Sarmiento, Hidalgo, Hilger, Hładczuk, Hobbs, Holland, Huckle, Jardine, Jasniewicz, Jean-Antoine~Piccolo, Jiménez-Arranz, Juaristi~Campillo, Julbe, Karbevska, Khanna, Kordopatis, Korn, Kóspál, Kostrzewa-Rutkowska, Kruszyńska, Kun, Laizeau, Lambert, Lanza, Lasne, Le~Campion, Lebreton, Lebzelter, Leccia, Lecoeur-Taibi, Liao, Licata, Lindstrøm, Lister,
  Livanou, Lobel, Lorca, Loup, Madrero~Pardo, Magdaleno~Romeo, Managau, Mann, Manteiga, Marchant, Marconi, Marcos, Marcos~Santos, Marín~Pina, Marinoni, Marocco, Marshall, Martin~Polo, Martín-Fleitas, Marton, Mary, Masip, Massari, Mastrobuono-Battisti, McMillan, Messina, Michalik, Millar, Mints, Molina, Molinaro, Molnár, Monari, Monguió, Montegriffo, Montero, Mor, Mora, Morbidelli, Morris, Muraveva, Murphy, Musella, Nagy, Noval, Ocaña, Ogden, Ordenovic, Osinde, Pagani, Pagano, Palaversa, Palicio, Pallas-Quintela, Panahi, Payne-Wardenaar, Peñalosa~Esteller, Penttilä, Pichon, Piersimoni, Pineau, Plachy, Plum, Poggio, Prša, Pulone, Racero, Ragaini, Rainer, Raiteri, Ramos, Ramos-Lerate, Regibo, Richards, Rios~Diaz, Ripepi, Riva, Rix, Rixon, Robichon, Robin, Robin, Roelens, Rogues, Rohrbasser, Romero-Gómez, Rowell, Royer, Ruz~Mieres, Rybicki, Sáez~Núñez, Sagristà~Sellés, Salguero, Samaras, Sanchez~Gimenez, Sanna, Santoveña, Sarasso, Schultheis, Sciacca, Segol, Segovia, Semeux, Siddiqui, Siebert,
  Siltala, Silvelo, Slezak, Slezak, Smart, Snaith, Solano, Solitro, Souami, Souchay, Spagna, Spina, Spoto, Steele, Steidelmüller, Stephenson, Süveges, Surdej, Szabados, Szegedi-Elek, Taris, Taylor, Teixeira, Tolomei, Tonello, Torra, Torra, Torralba~Elipe, Trabucchi, Tsounis, Turon, Ulla, Unger, Vaillant, van Dillen, van Reeven, Vanel, Vecchiato, Viala, Vicente, Voutsinas, Weiler, Wevers, Wyrzykowski, Yoldas, Yvard, Zhao, Zorec, \& Zucker}]{gaia_collaboration_gaia_2023-1}
{Gaia Collaboration}, Arenou, F., Babusiaux, C., {et~al.} 2023{\natexlab{b}}, Astronomy and Astrophysics, 674, A34, \dodoi{10.1051/0004-6361/202243782}

\bibitem[{{Gaia Collaboration} {et~al.}(2024){Gaia Collaboration}, Panuzzo, Mazeh, Arenou, Holl, Caffau, Jorissen, Babusiaux, Gavras, Sahlmann, Bastian, Wyrzykowski, Eyer, Leclerc, Bauchet, Bombrun, Mowlavi, Seabroke, Teyssier, Balbinot, Helmi, Brown, Vallenari, Prusti, de~Bruijne, Barbier, Biermann, Creevey, Ducourant, Evans, Guerra, Hutton, Jordi, Klioner, Lammers, Lindegren, Luri, Mignard, Nicolas, Randich, Sartoretti, Smiljanic, Tanga, Walton, Aerts, Bailer-Jones, Cropper, Drimmel, Jansen, Katz, Lattanzi, Soubiran, Thévenin, van Leeuwen, Andrae, Audard, Bakker, Blomme, Castañeda, De~Angeli, Fabricius, Fouesneau, Frémat, Galluccio, Guerrier, Heiter, Masana, Messineo, Nienartowicz, Pailler, Riclet, Roux, Sordo, Gracia-Abril, Portell, Altmann, Benson, Berthier, Burgess, Busonero, Busso, Cacciari, Cánovas, Carrasco, Carry, Cellino, Cheek, Clementini, Damerdji, Davidson, de~Teodoro, Delchambre, Dell'Oro, Fraile~Garcia, Garabato, García-Lario, Haigron, Hambly, Harrison, Hatzidimitriou, Hernández,
  Hestroffer, Hodgkin, Jamal, Jevardat~de Fombelle, Jordan, Krone-Martins, Lanzafame, Löffler, Lorca, Marchal, Marrese, Moitinho, Muinonen, Nuñez~Campos, Oreshina-Slezak, Osborne, Pancino, Pauwels, Recio-Blanco, Riello, Rimoldini, Robin, Roegiers, Sarro, Schultheis, Smith, Sozzetti, Utrilla, van Leeuwen, Weingrill, Abbas, Ábrahám, Abreu~Aramburu, Ahmed, Altavilla, Álvarez, Anders, Anderson, Anglada~Varela, Antoja, Baig, Baines, Baker, Balaguer-Núñez, Balog, Barache, Barros, Barstow, Bartolomé, Bashi, Bassilana, Baudeau, Becciani, Bedin, Bellas-Velidis, Bellazzini, Beordo, Bernet, Bertolotto, Bertone, Bianchi, Binnenfeld, Blanco-Cuaresma, Bland-Hawthorn, Blazere, Boch, Bossini, Bouquillon, Bragaglia, Braine, Bratsolis, Breedt, Bressan, Brouillet, Brugaletta, Bucciarelli, Butkevich, Buzzi, Camut, Cancelliere, Cantat-Gaudin, Capilla~Guilarte, Carballo, Carlucci, Carnerero, Carretero, Carton, Casamiquela, Casey, Castellani, Castro-Ginard, Ceraj, Cesare, Charlot, Chaudet, Chemin, Chiavassa, Chornay,
  Chosson, Cooper, Cornez, Cowell, Crosta, Crowley, Cruz~Reyes, Dafonte, Dal~Ponte, David, de~Laverny, De~Luise, De~March, de~Torres, del Peloso, Delbo, Delgado, Delisle, Demouchy, Denis, Dharmawardena, Di~Giacomo, Diener, Distefano, Dolding, Dsilva, Enke, Fabre, Fabrizio, Faigler, Fatović, Fedorets, Fernández-Hernández, Fernique, Figueras, Fouron, Fragkoudi, Gai, Galinier, Garcia-Serrano, García-Torres, Garofalo, Gerlach, Geyer, Giacobbe, Gilmore, Girona, Giuffrida, Gomboc, Gomez, González-Santamaría, Gosset, Granvik, Gregori~Barrera, Gutiérrez-Sánchez, Haywood, Helmer, Hidalgo, Hilger, Hobbs, Hottier, Huckle, Jiménez-Arranz, Juaristi~Campillo, Kaczmarek, Kervella, Khanna, Kontizas, Kordopatis, Korn, Kóspál, Kostrzewa-Rutkowska, Kruszyńska, Kun, Lambert, Lanza, Lebreton, Lebzelter, Leccia, Lecoutre, Liao, Liberato, Licata, Livanou, Lobel, López-Miralles, Loup, Madarász, Mahy, Mann, Manteiga, Marcellino, Marchant, Marconi, Marín~Pina, Marinoni, Marshall, Martín~Lozano, Martin~Polo,
  Martín-Fleitas, Marton, Mascarenhas, Masip, Mastrobuono-Battisti, McMillan, Meichsner, Merc, Messina, Millar, Mints, Mohamed, Molina, Molinaro, Molnár, Monguió, Montegriffo, Monti, Mora, Morbidelli, Morris, Mudimadugula, Muraveva, Musella, Nagy, Nardetto, Navarrete, Oh, Ordenovic, Orenstein, Pagani, Pagano, Palaversa, Palicio, Pallas-Quintela, Pawlak, Penttilä, Pesciullesi, Pinamonti, Plachy, Planquart, Plum, Poggio, Pourbaix, Price-Whelan, Pulone, Rabin, Rainer, Raiteri, Ramos, Ramos-Lerate, Ratajczak, Re~Fiorentin, Regibo, Reylé, Ripepi, Riva, Rix, Rixon, Robert, Robichon, Robin, Romero-Gómez, Rowell, Ruz~Mieres, Rybicki, Sadowski, Sagristà~Sellés, Sanna, Santoveña, Sarasso, Sarmiento, Sarrate~Riera, Sciacca, Ségransan, Semczuk, Shahaf, Siebert, Slezak, Smart, Snaith, Solano, Solitro, Souami, Souchay, Spitoni, Spoto, Squillante, Steele, Steidelmüller, Surdej, Szabados, Taris, Taylor, Teixeira, Tepper-Garcia, Thuillot, Tolomei, Tonello, Torra, Torralba~Elipe, Trabucchi, Trentin, Tsantaki, Turon,
  Ulla, Unger, Valtchanov, Vanel, Vecchiato, Vicente, Villar, Weiler, Zhao, Zorec, Zucker, Župić, \& Zwitter}]{gaia_collaboration_discovery_2024}
{Gaia Collaboration}, Panuzzo, P., Mazeh, T., {et~al.} 2024, Astronomy and Astrophysics, 686, L2, \dodoi{10.1051/0004-6361/202449763}

\bibitem[{{Generozov} \& {Perets}(2024)}]{generozov_perets_2024}
{Generozov}, A., \& {Perets}, H.~B. 2024, \apj, 964, 83, \dodoi{10.3847/1538-4357/ad2356}

\bibitem[{{Giacobbo} {et~al.}(2018){Giacobbo}, {Mapelli}, \& {Spera}}]{giacobbo_2018}
{Giacobbo}, N., {Mapelli}, M., \& {Spera}, M. 2018, \mnras, 474, 2959, \dodoi{10.1093/mnras/stx2933}

\bibitem[{{Gilkis} \& {Mazeh}(2024)}]{gilkis_2024}
{Gilkis}, A., \& {Mazeh}, T. 2024, \mnras, 535, L44, \dodoi{10.1093/mnrasl/slae091}

\bibitem[{{Green} {et~al.}(2019){Green}, {Schlafly}, {Zucker}, {Speagle}, \& {Finkbeiner}}]{green_2019}
{Green}, G.~M., {Schlafly}, E., {Zucker}, C., {Speagle}, J.~S., \& {Finkbeiner}, D. 2019, \apj, 887, 93, \dodoi{10.3847/1538-4357/ab5362}

\bibitem[{{Hafen} {et~al.}(2022){Hafen}, {Stern}, {Bullock}, {Gurvich}, {Yu}, {Faucher-Gigu{\`e}re}, {Fielding}, {Angl{\'e}s-Alc{\'a}zar}, {Quataert}, {Wetzel}, {Starkenburg}, {Boylan-Kolchin}, {Moreno}, {Feldmann}, {El-Badry}, {Chan}, {Trapp}, {Kere{\v{s}}}, \& {Hopkins}}]{hafen_2022}
{Hafen}, Z., {Stern}, J., {Bullock}, J., {et~al.} 2022, \mnras, 514, 5056, \dodoi{10.1093/mnras/stac1603}

\bibitem[{Halbwachs {et~al.}(2023)Halbwachs, Pourbaix, Arenou, Galluccio, Guillout, Bauchet, Marchal, Sadowski, \& Teyssier}]{halbwachs_gaia_2023}
Halbwachs, J.-L., Pourbaix, D., Arenou, F., {et~al.} 2023, Astronomy and Astrophysics, 674, A9, \dodoi{10.1051/0004-6361/202243969}

\bibitem[{Hobbs {et~al.}(2005)Hobbs, Lorimer, Lyne, \& Kramer}]{hobbs_statistical_2005}
Hobbs, G., Lorimer, D.~R., Lyne, A.~G., \& Kramer, M. 2005, Monthly Notices of the Royal Astronomical Society, 360, 974, \dodoi{10.1111/j.1365-2966.2005.09087.x}

\bibitem[{{Hopkins} {et~al.}(2018){Hopkins}, {Wetzel}, {Kere{\v{s}}}, {Faucher-Gigu{\`e}re}, {Quataert}, {Boylan-Kolchin}, {Murray}, {Hayward}, {Garrison-Kimmel}, {Hummels}, {Feldmann}, {Torrey}, {Ma}, {Angl{\'e}s-Alc{\'a}zar}, {Su}, {Orr}, {Schmitz}, {Escala}, {Sanderson}, {Grudi{\'c}}, {Hafen}, {Kim}, {Fitts}, {Bullock}, {Wheeler}, {Chan}, {Elbert}, \& {Narayanan}}]{hopkins_2018}
{Hopkins}, P.~F., {Wetzel}, A., {Kere{\v{s}}}, D., {et~al.} 2018, \mnras, 480, 800, \dodoi{10.1093/mnras/sty1690}

\bibitem[{{Hurley} {et~al.}(2002){Hurley}, {Tout}, \& {Pols}}]{hurley_2002}
{Hurley}, J.~R., {Tout}, C.~A., \& {Pols}, O.~R. 2002, \mnras, 329, 897, \dodoi{10.1046/j.1365-8711.2002.05038.x}

\bibitem[{{Iorio} {et~al.}(2024){Iorio}, {Torniamenti}, {Mapelli}, {Dall'Amico}, {Trani}, {Rastello}, {Sgalletta}, {Rinaldi}, {Costa}, {Dahl-Lahtinen}, {Escobar}, {Korb}, {Vaccaro}, {Lacchin}, {Mestichelli}, {Di Carlo}, {Spera}, \& {Arca Sedda}}]{iorio_2024}
{Iorio}, G., {Torniamenti}, S., {Mapelli}, M., {et~al.} 2024, \aap, 690, A144, \dodoi{10.1051/0004-6361/202450531}

\bibitem[{Janssens {et~al.}(2022)Janssens, Shenar, Sana, Faigler, Langer, Marchant, Mazeh, Schürmann, \& Shahaf}]{janssens_uncovering_2022}
Janssens, S., Shenar, T., Sana, H., {et~al.} 2022, Astronomy and Astrophysics, 658, A129, \dodoi{10.1051/0004-6361/202141866}

\bibitem[{{Kotko} {et~al.}(2024){Kotko}, {Banerjee}, \& {Belczynski}}]{kotko_2024}
{Kotko}, I., {Banerjee}, S., \& {Belczynski}, K. 2024, \mnras, \dodoi{10.1093/mnras/stae2591}

\bibitem[{{Kroupa}(2001)}]{kroupa_2001}
{Kroupa}, P. 2001, \mnras, 322, 231, \dodoi{10.1046/j.1365-8711.2001.04022.x}

\bibitem[{{Kruckow} {et~al.}(2024){Kruckow}, {Andrews}, {Fragos}, {Holl}, {Bavera}, {Briel}, {Gossage}, {Kovlakas}, {Rocha}, {Sun}, {Srivastava}, {Xing}, \& {Zapartas}}]{kruckow_2024}
{Kruckow}, M.~U., {Andrews}, J.~J., {Fragos}, T., {et~al.} 2024, \aap, 692, A141, \dodoi{10.1051/0004-6361/202452356}

\bibitem[{{K{\"u}pper} {et~al.}(2011){K{\"u}pper}, {Maschberger}, {Kroupa}, \& {Baumgardt}}]{kupper_2011}
{K{\"u}pper}, A. H.~W., {Maschberger}, T., {Kroupa}, P., \& {Baumgardt}, H. 2011, \mnras, 417, 2300, \dodoi{10.1111/j.1365-2966.2011.19412.x}

\bibitem[{{Lada} \& {Lada}(2003)}]{lada_2003}
{Lada}, C.~J., \& {Lada}, E.~A. 2003, \araa, 41, 57, \dodoi{10.1146/annurev.astro.41.011802.094844}

\bibitem[{{Li} {et~al.}(2024){Li}, {Zhu}, {Lu}, {L{\"u}}, {Li}, {Liu}, {Guo}, \& {Yu}}]{li_2024}
{Li}, Z., {Zhu}, C., {Lu}, X., {et~al.} 2024, \apjl, 975, L8, \dodoi{10.3847/2041-8213/ad8653}

\bibitem[{{Mapelli} {et~al.}(2017){Mapelli}, {Giacobbo}, {Ripamonti}, \& {Spera}}]{mapelli_2017}
{Mapelli}, M., {Giacobbo}, N., {Ripamonti}, E., \& {Spera}, M. 2017, \mnras, 472, 2422, \dodoi{10.1093/mnras/stx2123}

\bibitem[{{Mar{\'\i}n Pina} {et~al.}(2024){Mar{\'\i}n Pina}, {Rastello}, {Gieles}, {Kremer}, {Fitzgerald}, \& {Rando Forastier}}]{pina_2024}
{Mar{\'\i}n Pina}, D., {Rastello}, S., {Gieles}, M., {et~al.} 2024, \aap, 688, L2, \dodoi{10.1051/0004-6361/202450460}

\bibitem[{{Marks} \& {Kroupa}(2012)}]{marks_kroupa_2012}
{Marks}, M., \& {Kroupa}, P. 2012, \aap, 543, A8, \dodoi{10.1051/0004-6361/201118231}

\bibitem[{{Marshall} {et~al.}(2006){Marshall}, {Robin}, {Reyl{\'e}}, {Schultheis}, \& {Picaud}}]{marshall_2006}
{Marshall}, D.~J., {Robin}, A.~C., {Reyl{\'e}}, C., {Schultheis}, M., \& {Picaud}, S. 2006, \aap, 453, 635, \dodoi{10.1051/0004-6361:20053842}

\bibitem[{Mashian \& Loeb(2017)}]{mashian_hunting_2017}
Mashian, N., \& Loeb, A. 2017, Monthly Notices of the Royal Astronomical Society, 470, 2611, \dodoi{10.1093/mnras/stx1410}

\bibitem[{Nagarajan {et~al.}(2024)Nagarajan, El-Badry, Triaud, Baycroft, Latham, Bieryla, Buchhave, Rix, Quataert, Howard, Isaacson, \& Hobson}]{nagarajan_espresso_2024}
Nagarajan, P., El-Badry, K., Triaud, A. H. M.~J., {et~al.} 2024, Publications of the Astronomical Society of the Pacific, 136, 014202, \dodoi{10.1088/1538-3873/ad1ba7}

\bibitem[{{Pourbaix} {et~al.}(2022){Pourbaix}, {Arenou}, {Gavras}, {Gosset}, {Halbwachs}, {Siopis}, {Sozzetti}, {Bauchet}, {Damerdji}, {Delchambre}, {Delisle}, {Giacobbe}, {Holl}, {Jorissen}, {Lattanzi}, {Leclerc}, {Morel}, {Sadowski}, {Sahlmann}, \& {Segransan}}]{Pourbaix2022}
{Pourbaix}, D., {Arenou}, F., {Gavras}, P., {et~al.} 2022, {Gaia DR3 documentation Chapter 7: Non-single stars}, Gaia DR3 documentation, European Space Agency; Gaia Data Processing and Analysis Consortium.

\bibitem[{{Rastello} {et~al.}(2023){Rastello}, {Iorio}, {Mapelli}, {Arca-Sedda}, {Di Carlo}, {Escobar}, {Shenar}, \& {Torniamenti}}]{rastello_2023}
{Rastello}, S., {Iorio}, G., {Mapelli}, M., {et~al.} 2023, \mnras, 526, 740, \dodoi{10.1093/mnras/stad2757}

\bibitem[{Sana {et~al.}(2012)Sana, de~Mink, de~Koter, Langer, Evans, Gieles, Gosset, Izzard, Le~Bouquin, \& Schneider}]{sana_binary_2012}
Sana, H., de~Mink, S.~E., de~Koter, A., {et~al.} 2012, Science, 337, 444, \dodoi{10.1126/science.1223344}

\bibitem[{{Sanderson} {et~al.}(2020){Sanderson}, {Wetzel}, {Loebman}, {Sharma}, {Hopkins}, {Garrison-Kimmel}, {Faucher-Gigu{\`e}re}, {Kere{\v{s}}}, \& {Quataert}}]{sanderson_2020}
{Sanderson}, R.~E., {Wetzel}, A., {Loebman}, S., {et~al.} 2020, \apjs, 246, 6, \dodoi{10.3847/1538-4365/ab5b9d}

\bibitem[{{Schlegel} {et~al.}(1998){Schlegel}, {Finkbeiner}, \& {Davis}}]{schlegel_1998}
{Schlegel}, D.~J., {Finkbeiner}, D.~P., \& {Davis}, M. 1998, \apj, 500, 525, \dodoi{10.1086/305772}

\bibitem[{{Shao} \& {Li}(2019)}]{shao_li_2019}
{Shao}, Y., \& {Li}, X.-D. 2019, \apj, 885, 151, \dodoi{10.3847/1538-4357/ab4816}

\bibitem[{{Shikauchi} {et~al.}(2020){Shikauchi}, {Kumamoto}, {Tanikawa}, \& {Fujii}}]{shikauchi_2020}
{Shikauchi}, M., {Kumamoto}, J., {Tanikawa}, A., \& {Fujii}, M.~S. 2020, \pasj, 72, 45, \dodoi{10.1093/pasj/psaa030}

\bibitem[{Shikauchi {et~al.}(2022)Shikauchi, Tanikawa, \& Kawanaka}]{shikauchi_detectability_2022}
Shikauchi, M., Tanikawa, A., \& Kawanaka, N. 2022, The Astrophysical Journal, 928, 13, \dodoi{10.3847/1538-4357/ac5329}

\bibitem[{{Sigurdsson} \& {Phinney}(1993)}]{Sigurdsson1993}
{Sigurdsson}, S., \& {Phinney}, E.~S. 1993, \apj, 415, 631, \dodoi{10.1086/173190}

\bibitem[{{Tanikawa} {et~al.}(2024{\natexlab{a}}){Tanikawa}, {Cary}, {Shikauchi}, {Wang}, \& {Fujii}}]{tanikawa_2024}
{Tanikawa}, A., {Cary}, S., {Shikauchi}, M., {Wang}, L., \& {Fujii}, M.~S. 2024{\natexlab{a}}, \mnras, 527, 4031, \dodoi{10.1093/mnras/stad3294}

\bibitem[{{Tanikawa} {et~al.}(2023){Tanikawa}, {Hattori}, {Kawanaka}, {Kinugawa}, {Shikauchi}, \& {Tsuna}}]{tanikawa_2023}
{Tanikawa}, A., {Hattori}, K., {Kawanaka}, N., {et~al.} 2023, \apj, 946, 79, \dodoi{10.3847/1538-4357/acbf36}

\bibitem[{{Tanikawa} {et~al.}(2024{\natexlab{b}}){Tanikawa}, {Wang}, \& {Fujii}}]{tanikawa_bbh_2024}
{Tanikawa}, A., {Wang}, L., \& {Fujii}, M.~S. 2024{\natexlab{b}}, arXiv e-prints, arXiv:2407.03662, \dodoi{10.48550/arXiv.2407.03662}

\bibitem[{{Wang} {et~al.}(2015){Wang}, {Spurzem}, {Aarseth}, {Nitadori}, {Berczik}, {Kouwenhoven}, \& {Naab}}]{wang_2015}
{Wang}, L., {Spurzem}, R., {Aarseth}, S., {et~al.} 2015, \mnras, 450, 4070, \dodoi{10.1093/mnras/stv817}

\bibitem[{{Wang} {et~al.}(2022){Wang}, {Liao}, {Giacobbo}, {Olejak}, {Gao}, \& {Liu}}]{wang_2022}
{Wang}, Y., {Liao}, S., {Giacobbo}, N., {et~al.} 2022, \aap, 665, A111, \dodoi{10.1051/0004-6361/202243684}

\bibitem[{{Wetzel} {et~al.}(2016){Wetzel}, {Hopkins}, {Kim}, {Faucher-Gigu{\`e}re}, {Kere{\v{s}}}, \& {Quataert}}]{wetzel_2016}
{Wetzel}, A.~R., {Hopkins}, P.~F., {Kim}, J.-h., {et~al.} 2016, \apjl, 827, L23, \dodoi{10.3847/2041-8205/827/2/L23}

\bibitem[{Wiktorowicz {et~al.}(2020)Wiktorowicz, Lu, Wyrzykowski, Zhang, Liu, Justham, \& Belczynski}]{wiktorowicz_noninteracting_2020}
Wiktorowicz, G., Lu, Y., Wyrzykowski, L., {et~al.} 2020, The Astrophysical Journal, 905, 134, \dodoi{10.3847/1538-4357/abc699}

\bibitem[{{Yalinewich} {et~al.}(2018){Yalinewich}, {Beniamini}, {Hotokezaka}, \& {Zhu}}]{yalinewich_2018}
{Yalinewich}, A., {Beniamini}, P., {Hotokezaka}, K., \& {Zhu}, W. 2018, \mnras, 481, 930, \dodoi{10.1093/mnras/sty2327}

\bibitem[{{Yamaguchi} {et~al.}(2018){Yamaguchi}, {Kawanaka}, {Bulik}, \& {Piran}}]{yamaguchi_2018}
{Yamaguchi}, M.~S., {Kawanaka}, N., {Bulik}, T., \& {Piran}, T. 2018, \apj, 861, 21, \dodoi{10.3847/1538-4357/aac5ec}

\end{thebibliography}
\bibliographystyle{aasjournal}



\end{document}